\documentclass{article}
\usepackage{graphicx} 
\usepackage{amsmath} 
\usepackage{amssymb} 
\usepackage{color}
\usepackage{mathrsfs}
\usepackage{verbatim}

\newcommand{\bbibitem}{\bibitem}

\newcommand{\llabel}[1]{{\label{#1}}}

\newcommand{\ffoot}[1]{}

\renewcommand{\r}[1]{(\ref{#1})}
\newcommand{\bi}{\begin{itemize}}
\newcommand{\ei}{\end{itemize}}
\newcommand{\bd}{\begin{description}}
\newcommand{\ed}{\end{description}}
\newcommand{\be}{\begin{enumerate}}
\newcommand{\ee}{\end{enumerate}}

\renewcommand{\i}{\item}

\newcommand{\bqn}{\begin{eqnarray}}
\newcommand{\eqn}{\end{eqnarray}}
\newcommand{\eqnn}{\nonumber\end{eqnarray}}
\newcommand{\eqnl}[1]{\llabel{#1}\end{eqnarray}}

\newcommand{\nn}{\nonumber}
\newcommand{\noi}{\noindent}
\newcommand{\ba}[1]{\begin{array}{#1}}
\newcommand{\ea}{\end{array}}

\newcommand{\R}{\mathbb{R}}
\newcommand{\C}{\mathbb{C}}
\newcommand{\N}{\mathbb{N}}

\newcommand{\fine}{\end{document}}

\def \trait (#1) (#2) (#3){\vrule width #1pt height #2pt depth #3pt}
\def \qed{\hfill
        \trait (0.1) (6) (0)
        \trait (6) (0.1) (0)
        \kern-6pt   
        \trait (6) (6) (-5.9)
        \trait (0.1) (6) (0)
\medskip}
\def \qedmio{\hfill
             \trait (8) (8) (-0.1)
             \medskip}
\def \quadp{{\Huge $\qedmio$}}

\newtheorem{ml}{\bf Lemma}
\newtheorem{Theorem}{\bf Theorem}

\newtheorem{mrem}{\bf \underline{{\sl Remark}}}
\newtheorem{mcc}{\bf Corollary}
\newtheorem{Definition}{\bf Definition}
\newtheorem{mpr}{\bf Proposition}
\newtheorem{mproperty}{\bf Property}

\newcommand{\bt}{\begin{Theorem}}
\newcommand{\et}{\end{Theorem}}
\newcommand{\bl}{\begin{ml}}
\newcommand{\el}{\end{ml}}
\newcommand{\bp}{\begin{mpr}}
\newcommand{\ep}{\end{mpr}}
\newcommand{\bc}{\begin{mcc}}

\newcommand{\bproperty}{\begin{mproperty}}
\newcommand{\eproperty}{\end{mproperty}}

\newcommand{\ec}{\end{mcc}}
\newcommand{\bdeff}{\begin{Definition}}
\newcommand{\edeff}{\end{Definition}}
\newcommand{\brem}{\begin{mrem}\rm}
\newcommand{\erem}{\end{mrem}}

\newcommand{\ppotR}[3]
{

\begin{figure}\begin{center}
~\includegraphics[width=#3truecm]{./#1.eps}\\
\caption{#2}
\llabel{#1}
\end{center}
\end{figure}
\noindent$\!\!$}
       
\newcommand{\lam}{\lambda}
\newcommand{\g}{\gamma}
\newcommand{\al}{\alpha}
\newcommand{\eps}{\varepsilon}

\newcommand{\om}{\omega}

\renewcommand{\th}{\theta}

\newcommand{\con}{{\cal C}}

\newcommand{\sceq}{Schr\"{o}dinger equation\ }  
  
\newcommand{\neigh}{neighborhood }

\newcommand{\e}{\mbox{e}}

\newcommand{\und}{\underline}
\newcommand{\da}{\Delta_A^{-1}(0)}

\newcommand{\vep}{\varepsilon}

\newcommand{\cc}{ constant control }

\newcommand{\db}{\Delta_{B}^{-1}(0)}

\newcommand{\reachT}{{\cal R}(T)}

\newcommand{\ga}{\gamma}

\newcommand{\vect}[3]{\left(\begin{array}{ccc}{#1}\\{#2}\\{#3}\end{array}\right)}

\newcommand{\SB}{\mbox{{\bf S}}_{B}}
\newcommand{\OC}{{\mathbf{\Omega}\!\!\!\!\!\Omega}}

\newcommand{\GY}{  \mbox{{\LARGE $\g\!\!\!\g$}$_{\bar y}$}    } 
\newcommand{\ON}{\Omega_{\mbox{\footnotesize nasty}}}

\newcommand{\rr}{\mbox{{\tt {R}}}}

\begin{document} 

\begin{center} \noindent
{\LARGE{\sl{\bf Time Minimal Trajectories for a Spin $1/2$ Particle in a 
Magnetic field}}}
\vskip 1cm
Ugo Boscain, Paolo Mason\\
{\footnotesize SISSA, via Beirut 2-4 34014 Trieste, Italy}\\
e-mails {\tt boscain@sissa.it, mason@sissa.it}

\end{center}

\vspace{.5cm} \noindent \rm

\begin{quotation}
\noindent  {\bf Abstract}

In this paper we consider the minimum time population transfer problem for the $z$-component of the 
spin of a (spin 1/2) particle driven by a magnetic field,  controlled 
along the $x$ axis, with bounded amplitude. On the Bloch sphere (i.e. after a suitable Hopf 
projection), this problem can be attacked with techniques of optimal syntheses on 2-D 
manifolds.

Let $(-E,E)$ be the two energy levels, and $|\Omega(t)|\leq M$ the bound on the
field amplitude.
For each couple of values $E$ and $M$, we determine the time optimal 
synthesis starting from the level $-E$ 
and we provide the 
explicit
expression of the time optimal trajectories steering the state one to the
state two, in terms of a parameter that can be computed solving 
numerically 
a suitable equation. 

For $M/E<<1$, every time optimal trajectory is bang-bang and in particular the
corresponding control is periodic with frequency of the order of the 
resonance frequency $\om_R=2E$.

On the other side, for $M/E>1$, the time optimal trajectory steering 
the state one to the state two is 
bang-bang with exactly one switching. Fixed $E$ we also prove that for 
$M\to\infty$ the time needed to reach the state two tends to zero. 
In the case  $M/E>1$ there are time optimal trajectories containing a 
singular arc.

Finally we compare these results with some known results 
of Khaneja, Brockett and
Glaser and with those  obtained by controlling the magnetic field both on the $x$ and $y$ 
directions (or with one external field, but in the rotating wave 
approximation).

As byproduct we prove that the qualitative shape of the time optimal synthesis presents 
different  patterns, that cyclically 
alternate as $M/E\to0$, giving a partial proof of a conjecture formulated in a previous paper.

\end{quotation}


\vskip 1cm
\begin{center}
PREPRINT SISSA 82/2005/M 
\end{center}

\ffoot{
{\tt PAOLO CONTROLLA $P_N$ e $P_S$ $F_S$ !!!!!!!!!}\\
{\tt PAOLO CONTROLLA $[a,b[~~[a,b)$}\\
}
\newpage
\section{Introduction}
\subsection{Preliminaries}
The issue of designing an efficient transfer of population between
different atomic or molecular levels is crucial in atomic and molecular 
physics
(see e.g. \cite{bts,car1,dijon1,shorebook}). 
In the experiments, excitation and
ionization are often induced by means of a sequence of laser pulses.
The transfer should 
be as efficient as possible  in order to
minimize the effects of relaxation or decoherence that are always present.
In the recent past years, people started to approach the design of laser
pulses by using Geometric Control Techniques (see for instance
\cite{q1,rabitz,daless,brokko,rama}).
Finite dimensional closed quantum systems are in fact left (or
right) invariant control systems on $SU(n)$, or on the
corresponding
Hilbert sphere $S^{2n-1}\subset\C^n$, where $n$ is the number of atomic or
molecular  levels. For these kinds of systems very powerful techniques
were
developed  both for what concerns controllability
\cite{G-a,G-gb,G-jk,yuri}
and
optimal control \cite{agra-book,libro,jurd-book}. 
\\\\
The dynamics of a $n$-level quantum system is governed by the time 
dependent Schr\"odinger equation (in
a system of units such that $\hbar=1$),
\bqn
i \dot x(t)=(H_0+\sum_{j=1}^m \Omega_j(t) H_j)x(t)
\eqnl{eq-1}
where $x(.)$, defined on $[0,T]$ is a function taking values on the 
\underline{state space} which is $SU(n)$ 
(if we formulate the problem for time evolution operator) 
or the sphere $S^{2n-1}$ (if we formulate the problem for the  wave 
function).
The quantity $H_0$ called the 
\underline{drift Hamiltonian} is an Hermitian matrix, that is natural to 
assume diagonalized, i.e., $H_0=diag(E_1,...,E_n)$,
where $E_1,...,E_n$ are real numbers representing
the \underline{energy levels}. With no loss of generality we can assume 
$\sum_{j=1}^{n}E_j=0$.
The real valued \underline{controls} 
$\Omega_1(.),...,\Omega_m(.)$,  
represent the \underline{external pulsed
field}, while the matrices $H_j$ ($j=1,...,m$) are Hermitian 
matrices describing the coupling between the external fields 
and the system.
The time dependent Hamiltonian $H(t):=H_0+\sum_{j=1}^m \Omega_j(t) H_j$ is 
called the \underline{controlled Hamiltonian}.

The first problem that usually one would like to solve is 
the \underline{controllability problem}, i.e. proving that for every 
couple of points 
in the state space one can find controls steering the system from one 
point to the other. For applications, the most interesting initial and 
final states are of course the \underline{eigenstates of $H_0$}.

Thanks to the fact that the control system 
\r{eq-1} is a left invariant control system on the compact Lie group 
$SU(n)$,  
this happens if and only if
\bqn
\llabel{eq-controllability}
\mbox{Lie}\{iH_0,iH_1,...iH_m\}=su(n),
\eqn
(see for 
instance \cite{yuri}).  The problem of finding easily verifyable conditions 
under which \r{eq-controllability} is satisfied has been deeply studied in 
the literature (see for instance \cite{altafini}). Here we just recall that the 
condition \r{eq-controllability} is generic in the space of Hermitian 
matrices.\\\\
Once that controllability is proved one would like to steer the system, between 
two fixed points in the state space, in the most efficient way. Typical 
costs that are interesting to minimize for applications are:
\bi
\i {\bf Energy transfered by the controls to the system.}
$\displaystyle \int_0^T\sum_{j=1}^m \Omega_j^2(t) ~dt,$
\i {\bf Time of transfer.} In this case one can attack two different 
problems 
one with \und{bounded} and 
one with \underline{unbounded} controls.
\ei
The problem of minimizing time with unbounded controls is now well 
understood \cite{agra-chambrion,brokko}.
On the other side, the problems of minimizing energy, or time with 
bounded controls 
are hopeless in general. Indeed optimal trajectories 
must satisfy a first order necessary condition for optimality called the  
\und{Pontryagin Maximum Principle} (in the following PMP, 
see for instance \cite{agra-book,jurd-book}) that generalizes 
the Weierstra\ss \ conditions of Calculus of Variations to problems with 
non-holonomic constraints. For each optimal trajectory, the PMP
provides a lift to the cotangent bundle that is a solution to a suitable
pseudo--Hamiltonian system.
Hence, the first difficulty comes from  the problem of integrability of  
a Hamiltonian system (that generically is
not  integrable except for very special costs). Second, one should
manage
with some special solutions  of the PMP, the so called
\underline{abnormal} and \und{singular 
extremals} (see for instance \cite{bonnard-book}). Finally,  even if one 
is 
able to find all the solutions of the PMP (called \und{extremal 
trajectories}), it remains the problem of \underline{selecting}, among 
them, 
the \underline{optimal trajectories}, that usually is even a more 
difficult 
problem. For that purpose high order necessary conditions for optimality 
have been studied, like Clebsch-Legendre conditions, higher-order maximum 
principle, envelopes, conjugate points, index theory (cf. for instance 
\cite{agra-book,libro} and references therein).
For these reasons, usually, one can hope to find  a complete solution to
an optimal control problem   for very special
costs, dynamics and in low dimension only 
(\cite{agra-mario,libro,quattro,sch1}). 

In \cite{q1,q2,q3,q4} a special class of systems, 
for which the analysis can be pushed much further, was studied, namely 
systems such that  the drift term $H_0$ disappear in the interaction 
picture (by a unitary change of coordinates and a change
of controls). For these systems the controlled Hamiltonian reads
\bqn
H(t)&=&
\left(
\begin{array}{ccccc}
E_{1} & \mu_1\OC_{1}(t) & 0 & \cdots  & 0 \\
\mu_1\OC_{1}^{\ast}(t) & E_{2} & \mu_2\OC_{2}(t) & \ddots  & \vdots  
\\
0 & \mu_2\OC_{2}^{\ast}(t) & \ddots  & \ddots  & 0 \\
\vdots  & \ddots  & \ddots  & E_{n-1} & \mu_{n-1}\OC_{n-1}(t) \\
0 & \cdots  & 0 & \mu_{n-1}\OC_{n-1}^{\ast}(t) & E_{n}
\end{array}
\right)
\eqnl{eq-n-level}
Here $(^\ast)$ denotes the complex conjugation involution. The controls 
$\OC_1,\ldots,\OC_{n-1}$ are complex (they play the role of the real 
controls 
$\Omega_1,\ldots,\Omega_m$ in \r{eq-1} with $m=2(n-1)$)
and  $\mu_{j}>0,$ $(j=1,\ldots,n-1)$ are real constants describing the 
couplings (intrinsic to the quantum system) that we have restricted to
couple only levels $j$ and $j+1$ by pairs.
\ffoot{questa frase e' fuorviante: Notice that, thanks to possibility of eliminating the drift, 
for systems of type 
\r{eq-n-level} the problem of  minimizing energy is equivalent (up to reparametrization) to the 
problem of minimizing time with the constraint
$\sum_{j=1}^{n-1} |\OC_j(t)  |^2\leq1$.}

For $n=2$ the dynamics \r{eq-n-level} describes the evolution of the $z$ 
component of the spin of a (spin 1/2) particle  
driven by a magnetic field controlled both along the $x$ and 
$y$ axes, while for $n\geq2$ it represents the first $n$ levels of the 
spectrum of a 
molecule in the \underline{rotating wave approximation} 
(see for instance \cite{allen}), and 
assuming that each external fields couples only close levels.  
The complete solution to the optimal 
control problem between eigenstates of $H_0=diag(E_1,\ldots,E_n)$, has been 
constructed for $n=2$ and 
$n=3$, for the minimum time problem with bounded controls (i.e.,
$|\OC_j|\leq M_j$) and for the  
minimum energy problem  $\int_0^T\sum_{j=1}^{n-1} |\OC_j(t)  |^2~dt$ (with fixed final time).

\brem
\llabel{r-rwa}
For the simplest case $n=2$  (studied in \cite{q1,daless}), 
the minimum time problem with bounded control and the minimum energy 
problem actually coincide. In this case the controlled Hamiltonian 
is
\bqn
H(t)=\left(\begin{array}{cc} -E& \OC(t) \\ \OC^\ast(t)
&E
\end{array}\right),~~~|\OC|\leq M,
\label{eq-hgcomplex}
\eqn
and the optimal 
trajectories, steering the system from the first to the second eigenstate 
of $H_0=diag(-E,E)$, correspond to controls in
\und{resonance} with the energy gap $2E$, and with maximal amplitude 
i.e. $\OC(t)=Me^{i[(2E)t+\phi]}$,
where $\phi\in[0,2\pi[$ is an arbitrary phase.
The quantity $\om_R=2E$ is called the \underline{resonance frequency}.
In this case, the time  of transfer $T_\C$ is proportional to 
the inverse of the laser amplitude. More precisely  (see for
instance \cite{q1}), 
$T_\C=\pi/(2 M).$
\erem

For $n=3$ the problem has been studied in \cite{q2,q4} and it is  much 
more complicated (in particular when the 
coupling constants $\mu_1$ and $\mu_2$ are different). In the case of 
minimum time with bounded controls, it requires some 
nontrivial technical 
tools of 2-D syntheses theory for distributional systems, that have been 
developed in \cite{q4}. 

For $n\geq4$ the problem is hopeless, but in \cite{q3} it has 
been proved that the optimal controls steering the system from any 
couple of eigenstates of $H_0$ are in \underline{resonance}, i.e. they  
oscillate with a frequency equal to the difference of energy between
the levels  that the  control is coupling. More precisely 
\bqn
\OC_j=A_j(t)e^{i[(E_{j+1}-E_{j})t +\phi_j]},~~~~{j=1,...,n-1}
\eqn 
where $A_j(.)$ 	are real functions describing the amplitude of the 
external fields and $\phi_j$ are arbitrary phases.
Actually, this result holds for more general systems, initial and 
final 
conditions, and costs (see \cite{q3}).\\\\
The problem of minimizing time with bounded controls or energy is 
even more difficult if it is not possible to eliminate the 
drift $H_0$. This happens,  for instance, for a system in the form 
\r{eq-n-level} with 
real controls $\OC_j(t)=\OC_j^\ast(t)$, $j=1,...,n-1$, as we are going to 
discuss now. (For more details 
on 
the elimination of the drift see \cite{q1,q2,q3}.)
\subsection{A spin 1/2 particle in a magnetic field}
In this paper we attack the simplest quantum 
mechanical model interesting 
for applications for which it is not possible to eliminate the drift, 
namely a \und{two-level
quantum system} driven by a \und{real control}. This 
system describes the evolution of the $z$
component of the spin of a (spin 1/2) particle
driven by a magnetic field controlled along the $x$ axis. Equivalently it 
describes 
the first $2$ levels of a 
molecule driven by an external field without the rotating wave approximation.
The dynamics is governed by the time dependent Schr\"odinger equation (in
a system of units such that $\hbar=1$):
\begin{eqnarray}
i\frac{d\psi(t)}{dt}=H(t)\psi(t),
\llabel{eq-se}
\end{eqnarray}
where $\psi(.)=(\psi_1(.),\psi_2(.))^T:[0,T]\to{{\C}}^2$,
$\sum_{j=1}^2|\psi_j(t)|^2=1$ (i.e. $\psi(t)$ belongs to the sphere
$S^3\subset {{\C}}^2$), and
\begin{eqnarray}
H(t)=\left(\begin{array}{cc} -E& \Omega(t) \\ \Omega(t)
&E
\end{array}\right),
\llabel{eq-hg1}
\end{eqnarray}
where $E>0$ and the control
$\Omega(.)$, is assumed  to be a real function. With the notation of formula \r{eq-1}, 
the drift Hamiltonian is  
$H_0={\footnotesize \left(\ba{cc}-E&0\\0&E\ea\right)}$, while
$H_1={\footnotesize \left(\ba{cc}0&1\\1&0\ea\right)}$,
and the 
controllability condition \r{eq-controllability} is satisfied.

Notice  that for a spin 1/2 system, 
it is equivalent to treat the problem 
for the wave function or for the time evolution operator since $S^3$ 
is diffeomorphic to $SU(2)$.
The aim is to induce a transition from the first eigenstate of $H_0$ 
(i.e., $|\psi_1|^2=1$) to any other \und{physical state}. 
We recall that two 
states $\psi,\psi'\in S^3$ are 
{physically equivalent} if they differ for a factor of phase.
More precisely by physical state we mean a point of the two dimensional 
sphere
(called the \und{Bloch sphere})  $\SB:=S^3/ \sim$ where
the equivalence relation $\sim$ is defined as follows: $\psi\sim\psi'$
(where $\psi,\psi'\in S^3$) if and only if  $\psi=\exp{(i\Phi)}\psi'$, for
some $\Phi\in[0,2\pi[$. The projection from $S^3$ to $\SB$ is called 
\und{Hopf projection} and it is given explicitly in the next section.
A particularly interesting transition is of course from the first to the 
second eigenstates of $H_0$ (i.e., from $|\psi_1|^2=1$  to $|\psi_2|^2=1$).

Due to the presence of the drift, in this case the  minimum time problem 
with bounded control and the minimum energy problem are different. 
In \cite{daless} the authors studied the minimum energy problem (in that 
case, optimal solutions can be expressed in terms of Elliptic functions),
while here we minimize
the time of transfer, with bounded field amplitude:
\begin{eqnarray}
|\Omega(t)|\leq M,~~~\mbox{for every $t\in[0,T]$},
\label{bound}
\end{eqnarray}
where $T$ is the time of the transition and $M>0$ 
represents the maximum amplitude
available. This problem requires completely different techniques with 
respect to those used in \cite{daless}.

Thanks to the reduction to a two dimensional problem (on the Bloch 
sphere), this problem
can be attacked with the techniques of optimal syntheses on 2-D manifolds 
developed  by Sussmann, Bressan, Piccoli and the first author, see for
instance \cite{ex-syn,quattro,due,sus2}
 and recently rewritten in
\cite{libro}. We make a brief recall of these techniques in 
Appendix \ref{a-synt}.

\subsection{The Control problem on the Bloch Sphere $\SB$}
An explicit  Hopf projection from $S^3$ to $\SB$ is given by:
\bqn
\Pi:\vect{\psi_1}{\psi_2}{\psi_3}\in
S^3\subset\C^2 \longmapsto y=\vect{y_1}{y_2}{y_3}=
\left(\begin{array}{c}
-2\,{\rm Re}(\psi_1^*\psi_2)\\
2\,{\rm Im}(\psi_1^*\psi_2)\\
|\psi_1|^2-|\psi_2|^2
\end{array}\right)
\in
\SB
\subset\R^3.
\eqn
Notice that $\Pi$ maps the first eigenstate of $H_0$ (i.e. $|\psi_1|^2=1$) 
to the \underline{north pole}  $P_N:=(0,0,1)^T$ of $\SB$,  and the second  
eigenstate 
(i.e. $|\psi_2|^2=1$)
to  the \underline{south pole}  $P_S:=(0,0,-1)^T$.

After setting $u(t)=\Omega(t)/M$,
 the \sceq\ \r{eq-se}, \r{eq-hg1} projects to the following single input 
affine system (clarified below, after normalizations),
\begin{eqnarray}
\dot{y}&=&F_S(y)+uG_S(y),~~|u|\leq1,\mbox{ ~~where:}\llabel{cs-p}\\
&&y\in\SB:=\{(y_1,y_2,y_3)\in\R^3,~~~\sum_{j=1}^3 
y_j^2=1\}\\
&&F_S(y):=k \cos(\al)\left(\begin{array}{c}
-y_2\\y_1\\0\end{array}\right),~~~~~
G_S(y):=k\sin(\al)\left(\begin{array}{c}
0\\-y_3\\y_2\end{array}\right),    \label{cs-FG}\\
&&\alpha:=\arctan \left( \frac{M}{E}\right)\in\,]0,\pi/2[,
~~~k:=2E/\cos(\al)=2\sqrt{M^2+E^2}.
\llabel{cs-u}
\end{eqnarray}
\brem 
{\bf (normalizations)} In the following, to simplify the notations, we
normalize $k=1$. This normalization corresponds to a reparametrization of 
the time.\ffoot{PAOLO scrivi $t\to....$} More precisely, if $T$ is the 
minimum time to steer the state 
$\tilde y$ to the state $\bar y$ for the system with $k=1$, the 
corresponding minimum time for the original system is $T/(2\sqrt{M^2+E^2})$. 
Sometimes we need also the original system \r{eq-se}, \r{eq-hg1} on $S^3$, 
with the normalization made in this remark, i.e. the system 
\ffoot{\r{eq-se}, \r{eq-hg1} with $t\to...$}
\begin{eqnarray}
i\frac{d\psi(t)}{dt}=\tilde{H}(t)\psi(t)\,,\quad \mbox{where}\quad\tilde{H}(t)=
\frac{1}{2}\sin\al\left(\begin{array}{cc} 
-\cot\al & u(t) \\ u(t) & \cot\al \end{array}\right).
\label{eq-norm}
\end{eqnarray}
We come back to the 
original
value of $k$ only in Section \ref{s-comparison}, where we compare our 
results with those of other authors.
\erem
We refer to Figure \ref{f-bloch}. The vector fields $F_S(y)$ and $G_S(y)$ 
(that play the role respectively of $H_0$ and $H_1$)
describe rotations respectively
around the axes $y_3$ and $y_1$. 
Let us define the vector fields corresponding to constant control $\pm1$,
\bqn
X^\pm_S(y):=F_S(y)\pm G_S(y). 
\eqn
The parameter $\al\in]0,\pi/2[$ (that is 
the only parameter of the problem) is the angle between the axes of 
rotations of $F_S$ and $X^+_S$. The case $\al\geq\pi/4$ (resp. 
$\al<\pi/4$) corresponds to $M\geq E$ (resp. $M<E$).  
\ppotR{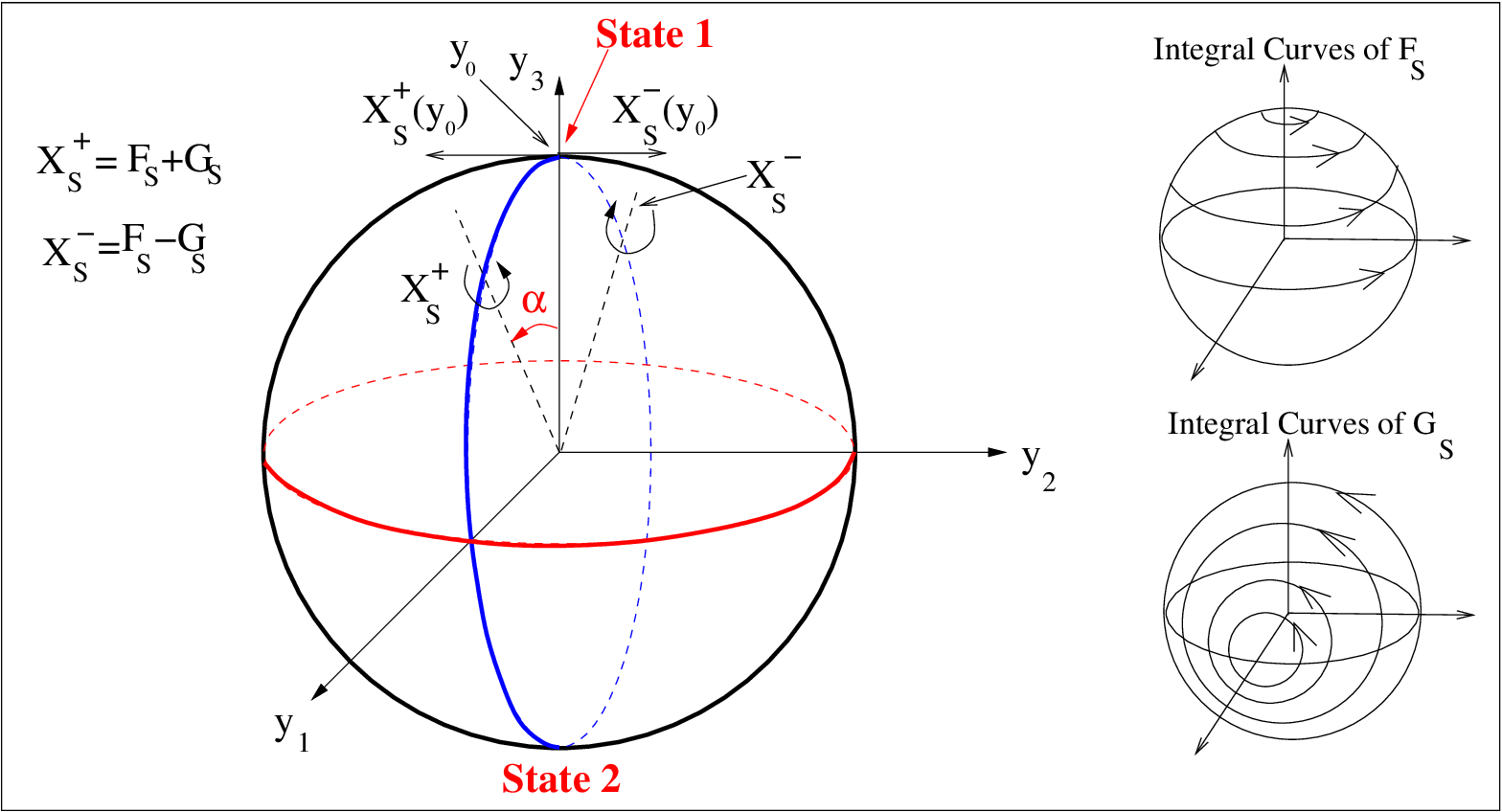}{The Bloch Sphere}{11}

~\vspace{-.9cm}
\bdeff
An admissible control $u(.)$ for the system \r{cs-p}--\r{cs-u} is a 
measurable function $u(.): [a,b]\to [-1,1]$,
while an admissible trajectory is a  Lipschitz functions 
$y(.):[a,b]\to\SB$ satisfying \r{cs-p} a.e.
for some  admissible control $u(.)$. 
If $y(.)$ is an admissible trajectory and $u(.)$ the corresponding 
control, we say that $(y(.),u(.))$ is an
admissible pair.
\edeff
For every $\bar y\in\SB$,  our minimization problem is then to find the 
admissible pair steering  the north pole to  $\bar y$ in minimum time.
More precisely\\\\ 
\medskip\noindent
{\bf Problem (P)} 
{\it Consider the control system
\r{cs-p}-\r{cs-u}. 
For every $\bar y\in\SB$, find an admissible pair 
$(y(.),u(.))$ defined on $[0,T]$ such that 
$y(0)=P_N$, $y(T)=\bar y$ and $y(.)$ is time optimal.
}\\\\
In Optimal Control the problem {\bf (P)} is known as the problem of 
computing the \underline{time optimal synthesis} for the system 
\r{cs-p}--\r{cs-u}. For more elaborated definitions of optimal synthesis 
see Appendix \ref{a-synt}, or \cite{libro,piccoli-sussmann} and 
references therein.
\bdeff {\bf (bang, singular for the problem  \r{cs-p}-\r{cs-u})}
\llabel{d-BS-1}
A control $u(.):[a,b]\to[-1,1]$  is said to be a \und{bang} control if 
$u(t)=+1$ 
a.e. in  $[a,b]$ or  $u(t)=-1$ a.e.  in  $[a,b]$. A control 
$u(.):[a,b]\to[-1,1]$ is said to be a 
\und{singular} control if $u(t)=0$, a.e. in  $[a,b]$. A finite concatenation 
of bang controls is 
called a \und{bang-bang} control. A \und{switching time} of $u(.)$ is a 
time 
$\bar t\in[a,b]$ such that, for every $\vep>0$, $u$ is not bang or 
singular on 
$(\bar t-\vep,\bar t+\vep)\cap 
[a,b]$.
A trajectory of the control system
\r{sys-gen} is said a bang
trajectory (or arc), singular trajectory (or arc), bang-bang 
trajectory, if it
corresponds respectively 
to a bang control, singular control, bang-bang control. If $\bar t$ is a switching time, the corresponding point on 
the trajectory $y(\bar t)$ is called a \und{switching point}.
\edeff
\brem
The definitions of singular trajectory and control, given above are very 
specific to our problem \r{cs-p}-\r{cs-u}. For the definition of 
singular trajectories for more  general systems see Definition 
\ref{d-BS-2}, 
Appendix \ref{a-pmp}.
\erem
In  \cite{y2} it was proved that, for the same problem  \r{cs-p}-\r{cs-u}, 
but in which  $y\in\R P^2$, for every couple of points there exists a time 
optimal trajectory joining them. Moreover it was proved that
every time optimal trajectory is a finite
concatenation of bang and singular trajectories. 
Repeating exactly the same arguments and recalling that $S^2$ is a 
double covering of $\R P^2$, one easily gets the same result on 
$\SB$. More precisely we have:
\bp
For the  problem
\r{cs-p}-\r{cs-u}, 
for each pair of points $p$ and $q$ belonging to $\SB$,
there exists a time optimal trajectory joining $p$ to $q$. 
Moreover every  time optimal trajectory for the 
problem \r{cs-p}-\r{cs-u} is a
finite concatenation of bang and singular trajectories.
\ep
Thus, the Fuller
phenomenon (i.e. existence of an optimal trajectory  joining two
points with an infinite
number of switchings in finite time)  never occurs. Notice that the previous proposition does not apply 
if $\al=0$ or $\al=\pi/2$, since in these cases the controllability property is lost.

\subsection{Purpose of the paper}
Our aim is to study problem {\bf (P)} for every possible value of the parameter
$\al$, giving a particular relief to the case in which $\bar y=P_S$ (i.e. 
to 
the optimal trajectory  steering the north to the south pole).

We will not be able to give a complete solution to the problem {\bf (P)}, 
without the help of numerical simulations. However, thanks to the theory 
developed in \cite{libro} we give a satisfactory description of the 
optimal synthesis. In the following we describe the main results and the structure of the paper.

For $\al<\pi/4$, every time optimal trajectory is bang-bang and in particular the
corresponding control is periodic, in the 
sense that for every fixed optimal trajectory  
the time between two consecutive switchings is constant. Moreover it tends to $\pi$ as $\al$ goes to 
$0$.
For  the original 
non normalized problem  this means  that for $M/E<<1$, the optimal 
control oscillates with frequency of the order of the 
resonance frequency $\om_R=2E$. In this case it is possible to
give a satisfactory description of the optimal synthesis excluding a 
neighborhood of the  south pole, in which we are able to compute the  optimal synthesis only numerically
(such results were already present in \cite{y2} as we see below).

On the other side, if $\al\geq\pi/4$ the computation of the optimal 
synthesis is simpler since the number of switchings needed to cover the whole sphere is small 
(less or equal than 2). In this case, for $\al$ big enough, we are also able to give the exact 
value of the time needed to cover the whole sphere. 
However, there is a new difficulty, namely the presence of  
singular arcs. Moreover the qualitative shape of the synthesis is 
rather different if $\al$ is close to $\pi/4$ or to $\pi/2$. 
A relevant fact is that this synthesis contains a singularity (the so called $(S,K)_3$) that is 
predicted by the general theory (see \cite{libro}, pag. 61 and 82), and was never observed out from 
{\sl ad hoc} examples. \ffoot{PAOLO bisogna decidere se dire che certi conti sono solo numerici}

The problem of  finding explicitly the optimal trajectories from the north pole $P_N$ 
to the south pole $P_S$, can be easily solved in the case $\al\geq\pi/4$ as a consequence of the 
construction of the time optimal synthesis. (Coming back to the original non 
normalized problem we also prove that fixed $E$, for $M\to\infty$ the time of transfer from $P_N$ 
to $P_S$ tends to zero.)

For $\al<\pi/4$ the problem is more 
complicated. 
However, using the symmetries of the problem, we are able to restrict the set  $\Xi$  of candidate optimal 
trajectories reaching the south pole, to a set containing at most  8 trajectories (half starting with 
control $+1$ and half starting with control $-1$, and switching exactly at the same times).
These trajectories are determined in terms of a parameter (the first switching time) that can be 
easily computed numerically solving suitable equations. Once  these trajectories are identified one 
can check by hands which are the optimal ones. 

\ffoot{qui volevo dire di piu', ma non ricordo esattamente cosa, forse qualcosa dal purpose of the paper}
The analysis can be pushed much forward. We also prove that the cardinality of $\Xi$ depends on the so 
called \und{normalized remainder} 
\bqn
\rr:=\frac{\pi}{2\al}-\left[\frac{\pi}{2\al}\right]\in[0,1[,
\eqnl{remainder} where 
$[~.~]$ denotes the integer part. In 
particular, for $\al$ small, we prove 
that if $\rr$ is close to zero then $\Xi$ contains exactly $8$ trajectories (and in particular there are
four optimal trajectories), while if  $\rr$ is
close to $1$  then $\Xi$ contains only $4$ trajectories (two of them are optimal).
The precise description of these facts is contained in Proposition \ref{alterna}. As a consequence, the 
qualitative shape of the time optimal synthesis presents 
different  patterns, that cyclically alternate, in the non controllability limit $\al\to0$, giving a 
partial proof 
of a conjecture formulated in a previous paper  (\cite{y2}), that was supported by numerical 
simulations, see Remark \ref{r-conj}.  This is probably the most interesting byproduct of this paper.

Finally we compare these results with some known results 
of Khaneja, Brockett and
Glaser and with those  obtained by controlling the magnetic field both on the $x$ and $y$ 
directions.

\medskip
The structure of the paper is as follows.
In Section \ref{hist} we briefly resume the results of paper \cite{y2} which are 
connected to our problem and the conjectures formulated therein. 
The main results
of the paper are described in Section \ref{s-main}, while the proofs are 
postponed
to Appendix \ref{a-pifferone}. In Appendix \ref{a-synt} we recall the 
main tools of the theory 
of optimal synthesis. In Appendix \ref{a-c} we determine the last
point reached by trajectories starting at $P_N$ and the time needed to
cover the whole sphere. 

\section{History of the problem and known facts}
\llabel{hist}

The problem {\bf (P)} (although with different
purposes)  was already partially studied
in \cite{y2}, in the case
$\al<\pi/4$. 
In that paper the aim was to 
give an estimate on the maximum number of switching
for time optimal trajectories on $SO(3)$ (problem first studied by 
Agrachev and Gamkrelidze in \cite{agra-sympl-x}, using index theory).

\begin{figure}
\begin{center}
\includegraphics[width=11truecm,angle=0]{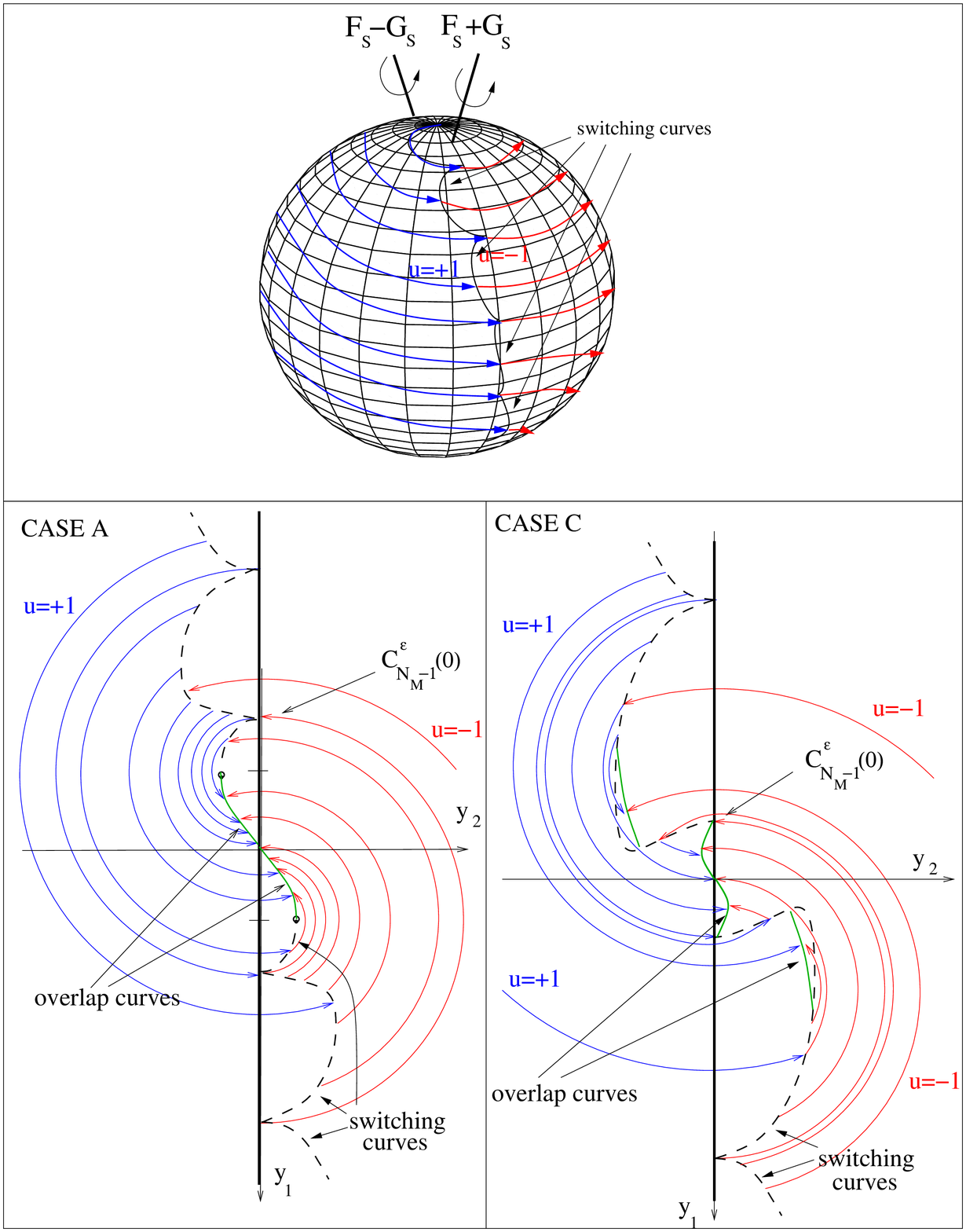} 
\caption{Synthesis on the sphere for $\al<\pi/4$ and conjectured shape in a neighborhood of the 
south pole}
\label{f-TUTTALASINTESI}
\end{center}
\end{figure}

In \cite{y2} it has been proved that, for the problem {\bf (P)} in the 
case $\al<\pi/4$, every optimal trajectory is bang-bang. 
More precisely,  it was proved that in 
the case $\al<\pi/4$, if $y(.)$ is a time optimal trajectory starting at 
the north pole, then it should satisfy the following properties:
\bd
\i[\bf i)] $y(.)$ is bang bang;
\i[{\bf ii)}] the 
duration $s_i$ of the first bang 
arc satisfies $s_i\in[0,\pi]$,
\i[{\bf iii)}] the time duration 
between two consecutive switchings is the same for all \underline{interior 
bang arcs} (i.e. excluding the first and the last bang) 
and it is the following function of $s_i$ defined in the interval $[0,\pi]$,
\bqn
v(s_i)=\pi+2 \arctan\left(\frac{\sin(s_i)}
{\cos(s_i)+\cot^2(\alpha)}\right).
\eqnl{v()}
One can immediately check that this function satisfies $v(0)=v(\pi)=\pi$ 
and $v(s_i)>\pi$ for every $s_i\in]0,\pi[$,
\i[{\bf iiii)}] the time duration of the last arc is $s_f\in[0,v(s_i)]$,
\ed
Properties {\bf i)}--{\bf iiii)} are illustrated in Figure \ref{f-sivs}.
Moreover, thanks to the analysis given in \cite{y2}, one easily get 
(always in the case $\al<\pi/4$):
\bd
\i[{\bf v)}] the number of switchings $N_y$ of $y(.)$ satisfies the 
following inequality  
\bqn
N_y\leq N_M:=\left[ \frac{\pi}{2\al}\right]+1
\eqnl{eq-nmax}
\ed
\ppotR{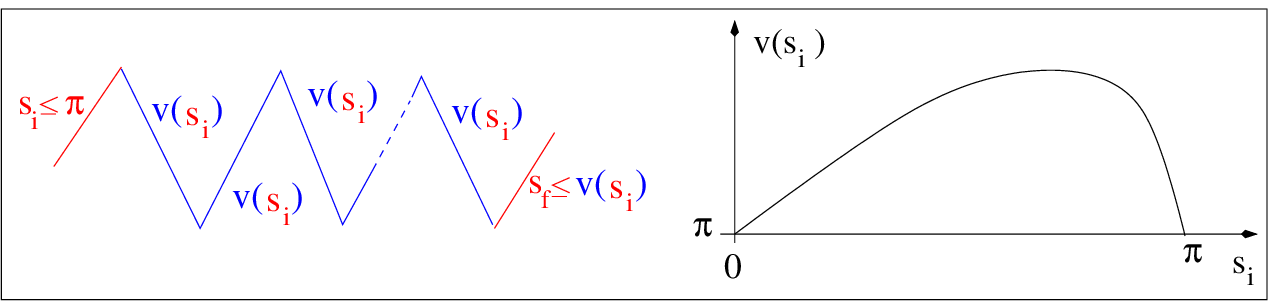}{Time optimal trajectories for $\al<\pi/4$}{12}
Conditions {\bf i)}-{\bf v)} define a set of 
\und{candidate optimal trajectories}.
The way in which these candidate optimal trajectories cover the whole 
sphere is shown in the top of Figure \ref{f-TUTTALASINTESI}. 

Consider the following curves, made by points where the
control switches from $+1$ to $-1$ or viceversa, 
called \und{switching
curves}, defined by induction
\bqn
C_1^{\eps}(s)= e^{X^{\eps}_Sv(s)}   e^{X^{-\eps}_Ss}  P_N,\ \ \
C^{\eps}_{k}(s)=e^{X^{\eps}_Sv(s)}C^{-\eps}_{k-1}(s),  \mbox{ (where 
$\eps=\pm1$ 
and $k=2,....,N_M-1$).}
\eqn
See the top of Figure \ref{f-TUTTALASINTESI}. 

Even if the analysis made in \cite{y2} was sufficient to the purpose of 
giving a bound on the maximum number of switchings for time optimal 
trajectories on $SO(3)$, some questions remained unsolved. In particular 
questions about 
\underline{local optimality} of 
the switching curves.
Roughly speaking we say that a switching 
curve is locally
optimal if it never ``reflects'' the
trajectories (see Figure \ref{f-nonlocop} A).\footnote{\llabel{foot-piede}
More precisely consider a smooth switching curve $C$ between two smooth vector field
$Y_1$ and $Y_2$ on a smooth two dimensional manifold.
Let $C(s)$ be a
smooth parametrization of $C$.
We say that $C$
is \underline{locally optimal} if, for every $s\in Dom(C)$, we have
$\dot C(s)\neq\al_1 Y_1(C(s))+\al_2 Y_2(C(s)), \mbox{ for every
}\al_1,\al_2\mbox{ s.t. }\al_1\al_2\geq0.$
The points of a switching curve on which this relation  is not
satisfied are usually called ``conjugate points''. See Figure \ref{f-nonlocop}.
}
When a family of trajectories is reflected by a switching curve then local 
optimality is lost and some \und{cut locus} appear in the optimal  
synthesis. 

\bdeff
A cut locus for the problem {\bf (P)} is a set of points reached at the 
same time by two (or more) optimal trajectories.  
A subset of a cut locus that is a 
connected $\con^1$ manifold is called 
\und{overlap curve}. 
\edeff

An example showing how a ``reflection'' 
on a switching curves generate a cut locus is portrayed in Figure 
\ref{f-nonlocop} B and C. More details are given later. 
In \cite{y2}, the following questions remain unsolved:
\ppotR{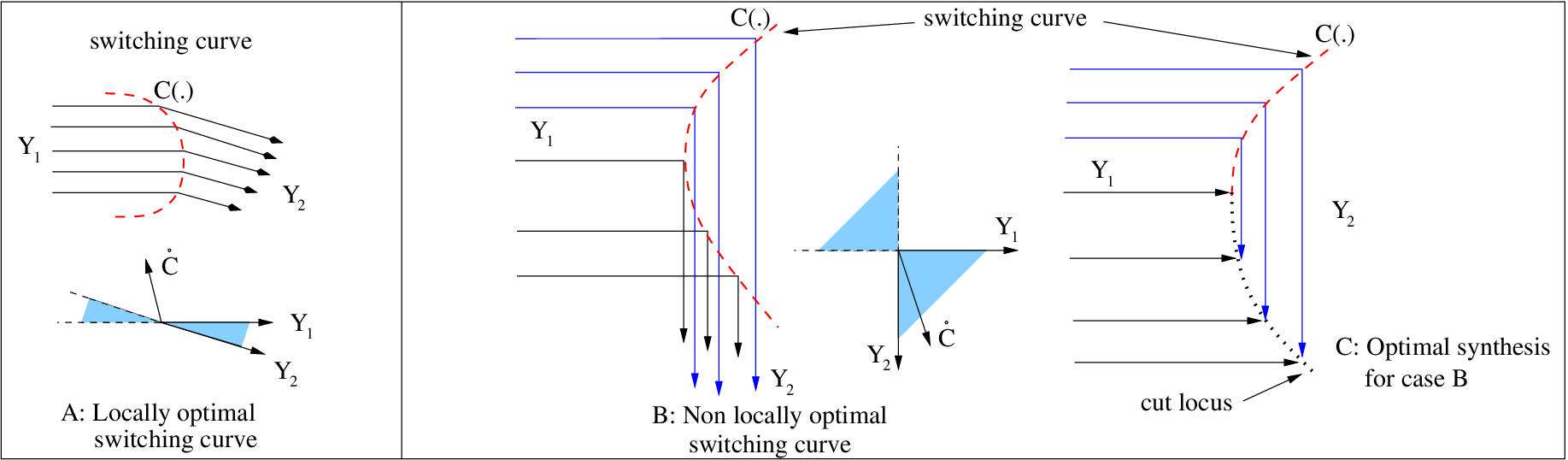}{Locally optimal switching curves and non locally optimal switching 
curves with the corresponding synthesis}{16.5}

\bd
\i[Question 1] Are the switching curves $C^\eps_k$, $k=1...,N_M-1$, 
locally optimal? More precisely, one would like to understand  how the 
candidate optimal trajectories described above are
going to lose optimality.
\ffoot{(i.e. if the loss of optimality is local or just global).}

\i[Question 2] What is the shape of the optimal synthesis in a \neigh of
the south pole?
\ed
Numerical simulations suggested some conjectures regarding 
the above questions. More precisely:
\bd
\i[C1] Define $k_{last}=\left[\frac{\pi-\al}{2\al}\right]-1$. 
Then the curves $C^\eps_k(s)$, ($k=1,...,N_M-1$) are locally optimal if 
and only if $k\leq k_{last}$. 
Notice that  $k_{last}\in\{N_M-3,N_M-2\}$.
\ed
Analyzing the evolution of the
minimum time wave front in a \neigh of the
south-pole, it is reasonable to conjecture that: 
\ffoot{vedere se mettere una ficura semplificata dei fronti tipo 
la f-s91.eps o anche se mettere una traiettoria con le estremali}
\bd
\i[C2] The shape of the optimal synthesis in a neighborhood of the 
south pole depends on the so called \und{remainder}\footnote{Notice that $r=2\al\rr$, where $\rr$ has 
been defined in Formula \r{remainder}. In conjecture {\bf C2}, we use the remainder $r$, to keep the 
same notation of \cite{y2}.} $r:=\pi-2\al \left[ \frac{\pi}{2\al}
\right].$
Notice that $r$ belongs to the interval $[0,2\al[$.
More precisely, we conjecture that for $\al\in]0,\pi/4[$, there 
exist two positive numbers $\al_1$ and $\al_2$ such that 
$0<\al_1<\al<\al_2<2\al$ and:
\bd
\i{}\underline{CASE A: $r\in]\al_2,2\al[$}.
The switching curve $C^\eps_{N_M-1}$ glues to an
overlap  curve that passes  through the origin (Fig.
\ref{f-TUTTALASINTESI}, Case ~A).

\i{}\underline{CASE B:  $r\in[\al_1,\al_2]$}. 
The switching curve $C^\eps_{N_M-1}$ is not reached by optimal 
trajectories in the interval $]0,\pi]$. At the point 
 $C^\eps_{N_M-1}(0)$ an overlap
curve starts and passes through the origin.

\i{}\underline{CASE C:  $r\in]0,\al_1[$}. The situation is
more complicated and it is
depicted in the bottom of Fig. \ref{f-TUTTALASINTESI}, Case~C.

\ed
\ed
{For $r=0$, the situation is the same as in CASE A, but for the
switching curve starting at $C^\eps_{N_M-2}(0)$.}

\section{Main Results}\llabel{s-main}
We give here a brief description of the main results of the paper. The corresponding
proofs are given in Appendix \ref{a-pifferone}. From now on we use the following
conventions.

\brem {\bf (notation)} The letter $B$ refers to a bang trajectory and the 
letter $S$ refers 
to a singular  trajectory. A concatenation of bang and singular trajectories
is labeled by the corresponding letter sequence, written in order from
left to right.
Sometimes, we use a subscript to indicate the time duration of a trajectory
so that we use $B_t$ to refer to a bang trajectory defined on an interval
of length $t$ and, similarly, $S_t$ for a singular trajectory defined on 
an interval of length $t$.
Moreover we indicate by $\ga^+$ (resp. $\ga^-$) the 
trajectory of \r{cs-p}--\r{cs-u} starting at the north pole at time zero
and corresponding to control $u\equiv 1$ (resp. $u\equiv 
-1$). Notice that $\ga^\pm$ are defined for every time, and are 
periodic. Finally we use the following subsets of $\SB$:  the circle of equation 
$y_3=0$ called  \und{equator},  the set $y_3>0$, called \und{north hemisphere} and the set
$y_3<0$, called 
\und{south hemisphere}. 
\llabel{r-4}
\erem
\subsection{Optimal synthesis for $\al\geq\pi/4$}

In this section we describe the time optimal synthesis for $\al\geq\pi/4$.
We divide $\SB$ in 8 open regions called 
$\Omega^\pm_1,...,\Omega^\pm_3$, 
$\ON^\pm$ 
and in 16 arcs (see 
Definition \ref{d-Avena}, and Figure \ref{f-farro}). 
For every point $\bar y\in\SB\setminus(\ON^+\cup\ON^-)$,  Theorem
\ref{t-Evena}  gives the optimal 
trajectories reaching $\bar y$. 

Unlike the $\al<\pi/4$ case, here it is possible to detect the presence of 
singular trajectories that
are optimal, and also of cut loci (even not only in a \neigh of the south 
pole).

The region $\ON^+$ (and similarly $\ON^-$) is more difficult to analyze. 
It contains a cut locus that should be 
determined numerically.  
Even if we  are not able to provide an  analytic
characterization of this locus, we are able to prove the following.
\bd
\i[i)] $\al=\arcsin(1/\sqrt[4]{2})$ is a bifurcation point for the optimal 
synthesis i.e.  the qualitative shape is different if 
$\al\in [\pi/4,\arcsin(1/\sqrt[4]{2})[$  (called {\bf Case 1}) or
$\al\in[\arcsin(1/\sqrt[4]{2}),\pi/2[$ (called {\bf Case 2}). 
More precisely, from the point $D^+:=\g^+(\pi)$, in {\bf Case 1} it starts an optimal  
switching curve, while in  {\bf Case 2} it starts an overlap 
curve (see Proposition~\ref{sw-over}).  The situation in $\ON^-$ is symmetric.

\i[ii)] The south pole belongs to 
the cut locus and it  is reached exactly by four optimal trajectories 
(see Proposition~\ref{propA}).

\ed

Numerical computations show that in  {\bf Case 2}, the cut locus in $\ON^+$ is an overlap curve 
connecting $D^+$ with the south pole, while 
in  {\bf Case 1}, the switching curve starting from  $D^+$ loses local optimality at a point of $\ON^+$
and connects to an overlap curve which reaches 
the south pole (see Figure \ref{f-sintesidoppia}). 
Remark \ref{r-nocal} explains that in {\bf Case 2} it is not necessary  to compute the cut locus lying 
in $\ON^+$ 
to get the expression of the optimal trajectory connecting $P_N$ to a point of  $\ON^+$.
The situation in $\ON^-$ is symmetric.\\\\
Let us start with the description of the optimal synthesis in $\SB\setminus(\ON^+\cup\ON^-)$.
Even if Definition \ref{d-Avena} and  Theorem \ref{t-Evena} look 
complicated, the shape of the 
optimal synthesis is quite simple as it is shown in Figure \ref{f-sintesidoppia}.
\ppotR{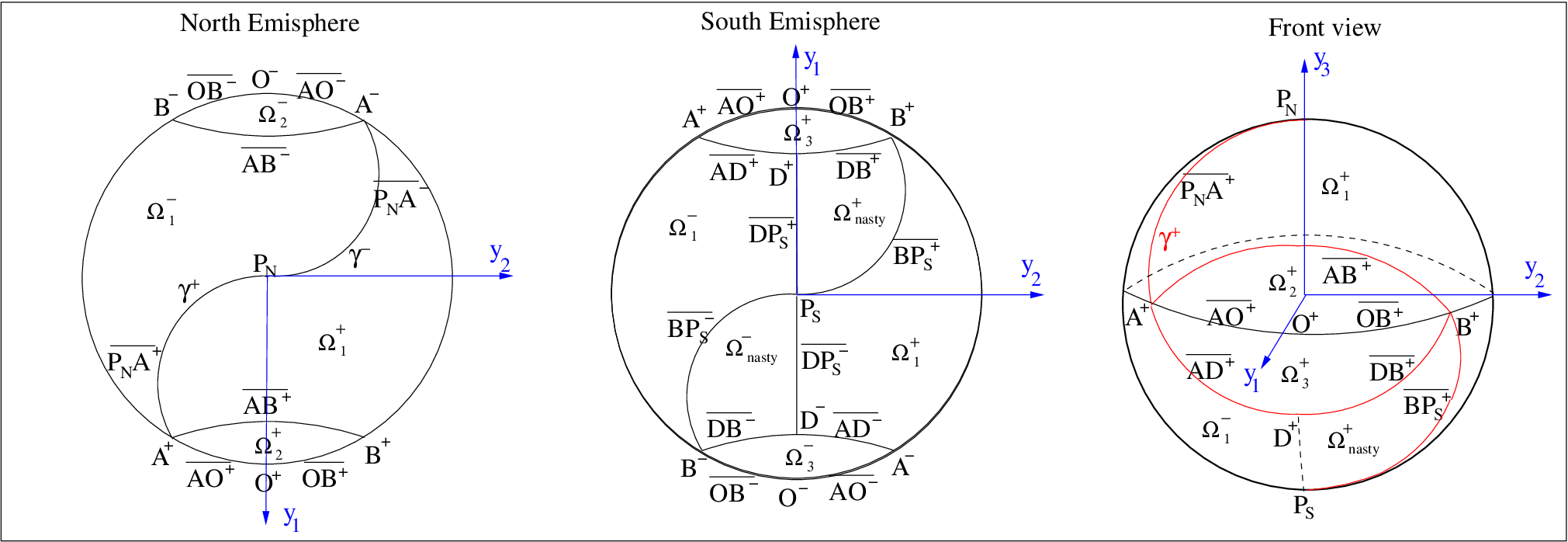}{Definition \ref{d-Avena}}{16}
\bdeff
\llabel{d-Avena} 
According to Figure \ref{f-farro}, let us define the following curves on 
$\SB$.

\bi
\i Let $t_1$ be the first time at which $\g^+$ intersects the equator 
and let $A^+:=\g^+(t_1)$ (notice that $t_1=\pi-\arccos(\cot^2(\al))$).
Define 
$\overline{P_NA^+}=Supp\left(\g^+|_{[0,t_1]}\right)$.   

\i Let $\xi^-$ be the trajectory corresponding to control $-1$, 
starting at time zero from $A^+$. Let $t_2$ be the first positive time at 
which  $\xi^-$ intersects the equator (notice that $t_2=2 
\arccos(\cot^2(\al))$).  Define 
$B^+:=\xi^-(t_2)$ and 
$\overline{AB^+}=Supp\left(\xi^-|_{[0,t_2]}\right)$.

\i Let $O^+=(1,0,0)$. Define 
$\overline{AO^+}$ (resp. $\overline{OB^+}$) as the support of the 
trajectory 
corresponding to control 
zero, starting at $A^+$  (resp. $O^+$) and ending at $O^+$ (resp. $B^+$).

\i Recall that  $D^+=\g^+(\pi)$, and define 
$\overline{AD^+}=Supp\left(\g^+|_{[t_1,\pi]}\right)$, 
$\overline{DB^+}=Supp\left(\g^+|_{[\pi,t_3]}\right)$, where  $t_3$ is 
the second intersection time of $\g^+$ with the equator 
(notice that $t_3=\pi+\arccos(\cot^2(\al))=t_1+t_2$).

\i Let $\overline{BP_S^+}$ the support of the trajectory corresponding to 
control $-1$, starting at $B^+$ and ending at the south pole.

\i Let $\overline{DP_S^+}$ the connected subset of the meridian $y_2=0$, 
lying in the south hemisphere and connecting the point $D^+$ to the south 
pole. 

\ei
Similarly define 
$A^-,$ $B^-,$ $O^-$, $D^-$, $\overline{P_NA^-}$, 
$\overline{AB^-}$, $\overline{AO^-}$, 
$\overline{OB^-}$, $\overline{AD^-}$,  $\overline{DB^-}$,  
$\overline{BP_S^-}$,  $\overline{DP_S^-}$ .

According to Figure \ref{f-farro} define 
$\Omega_1^\pm,\ldots,\Omega_4^\pm,\ON^\pm$ 
as the open connected components of 
the open set obtained subtracting from $\SB$  all the arcs 
defined above. 
\edeff
The following theorem holds for every $\al\in]\pi/4,\pi/2[$. For the particular
value $\al=\pi/4$ the claims of the theorem must be modified. Such changes are reported in
Remark~\ref{r-pigrecoquarti}. 

\bt
\llabel{t-Evena}
Let $\GY$ be the set of time optimal trajectories steering the north pole to 
$\bar y$. We have the 
following:
\bd
\i[T1.] If $\bar y\in \overline{P_NA^+}$ then $\GY$ is made by a unique 
trajectory corresponding 
to control $+1$ of the form $B_t$, with $t\leq t_1$.
\i[T2.] If $\bar y\in\overline{AB^+}\setminus B^+$ then $\GY$ is made by a 
unique trajectory of the form $B_{t_1}B_t$ (with the first bang corresponding to control $+1$).
\i[T3.] If $\bar y\in \overline{AO^+}$ then $\GY$  is made by a unique 
trajectory of the form 
$B_{t_1}S_s$ (with the first bang corresponding to control $+1$).

\i[T4.]  If $\bar y\in \overline{OB^+}\setminus O^+$ then $\GY$ is
made by two trajectories of the form $B_{t_1}S_s B_t$, both starting with 
control $+1$ and ending respectively with control $+1$ and $-1$. These 
two trajectories have the same values of $s\geq 0$
and $t>0$.

\i[T5.] If $\bar y\in\overline{AD^+}$ then $\GY$ is made by a unique 
trajectory
corresponding to control $+1$ of the form $B_t$, with $t\in[t_1,\pi]$.

\i[T6.] If $\bar y\in\overline{DB^+}\setminus B^+$ then $\GY$ is made by a 
unique trajectory
corresponding to control $+1$ of the form $B_t$, with $t\in[\pi,t_3[$.

\i[T7.] If $\bar y\in\overline{BP_S^+}$ then $\GY$ is made by two 
trajectories
respectively of the form $B_{t_1}B_t$ and $B_{t_3}B_{t-t_2}$ and starting 
with 
control $+1$.

\i[T8.] If $\bar y\in\Omega_1^+\cup(\overline{DP_S^-}\setminus P_S)$, then  
$\GY$  is made by a unique trajectory of the form 
$B_tB_{t'}$, with  $0\leq t<t_1$ and  the first bang corresponding to 
control $+1$.

\i[T9.] If  $\bar y\in\Omega_2^+$, then $\GY$ is made by a unique 
trajectory of 
the form $B_{t_1}S_sB_t$, with $s>0$, the first bang arc and the last bang 
arc
corresponding respectively to control $+1$ and $-1$.

\i[T10.] If $\bar y\in \Omega_3^+$, then $\GY$ is made by a unique 
trajectory of
the form $B_{t_1}S_sB_t$, with  $s>0$ and both bang arcs corresponding to 
control $+1$.

\i[T11.]  If $\bar y=P_S$  then  $\GY$  is made by the four trajectories 
of 
the 
form $B_{t_1}B_{t_3}$ and $B_{t_3}B_{t_1}$.

\i[T12.] If $\bar y\in\ON^+$ then every trajectory of $\GY$ is bang-bang 
with at most two switchings.

\ed
If $\bar y$ belongs to one of the remaining sets defined above, 
the description of the optimal strategy is analogous, by 
symmetry.
\et

\brem 
\llabel{r-pigrecoquarti}
In the case $\al=\pi/4$ some changes in the previous statement are required.
In particular the points $A^+$,$B^+$,$O^+$ and $D^+$ coincide (also the 
points  $A^-$,$B^-$,$O^-$ and $D^-$ coincide) and, consequently, there are 
no optimal trajectories containing singular arcs. Another immediate consequence 
of this fact is that there are only two optimal trajectories  reaching 
the south pole, of the form 
$B_\pi B_\pi$.
\erem

\brem
Notice that every point of $\overline{OB^+}\setminus O^+$,  
$\overline{OB^-}\setminus O^-$,  $\overline{BP_S^+}$,  
$\overline{BP_S^-}$ is reached by more than one optimal 
trajectory, i.e. it belongs to the cut locus. Other points of the cut 
locus can be identified numerically in $\ON^+$ and  $\ON^-$ as explained 
in the next section.
\erem

\brem \llabel{r-notall}
In Theorem \ref{t-Evena} we do not specify all the durations of the 
bang arcs. However the missing ones 
can be  obtained simply by following the switching strategy backwards. 
\erem

\brem
Note that the region reached by optimal trajectories containing a singular arc 
$\Omega_2^\pm\cup\Omega_3^\pm\cup \overline{AO^\pm}\cup  \overline{OB^\pm}$ become bigger and bigger as 
$\al$ tends to $\pi/2$. Moreover, in this limit, since the modulus of the 
drift $F_S$ becomes smaller and smaller, the time 
needed to cover such region tends to infinity. Notice however that the time needed to reach $P_S$ is 
always $2\pi$. The time needed to reach every point of the sphere for $\al$ big enough, and the last 
point reached by an optimal trajectory containing a singular arc, can be computed explicitly. This is 
done in Appendix \ref{a-c}.
\erem

Since the case  $\bar y=P_S$ is 
important also for the determination of the cut locus in  
$\ON^+\cup\ON^-$, it is reported in the next section as a 
separate proposition (see Proposition \ref{propA}).

\begin{figure}
\begin{center}
\input{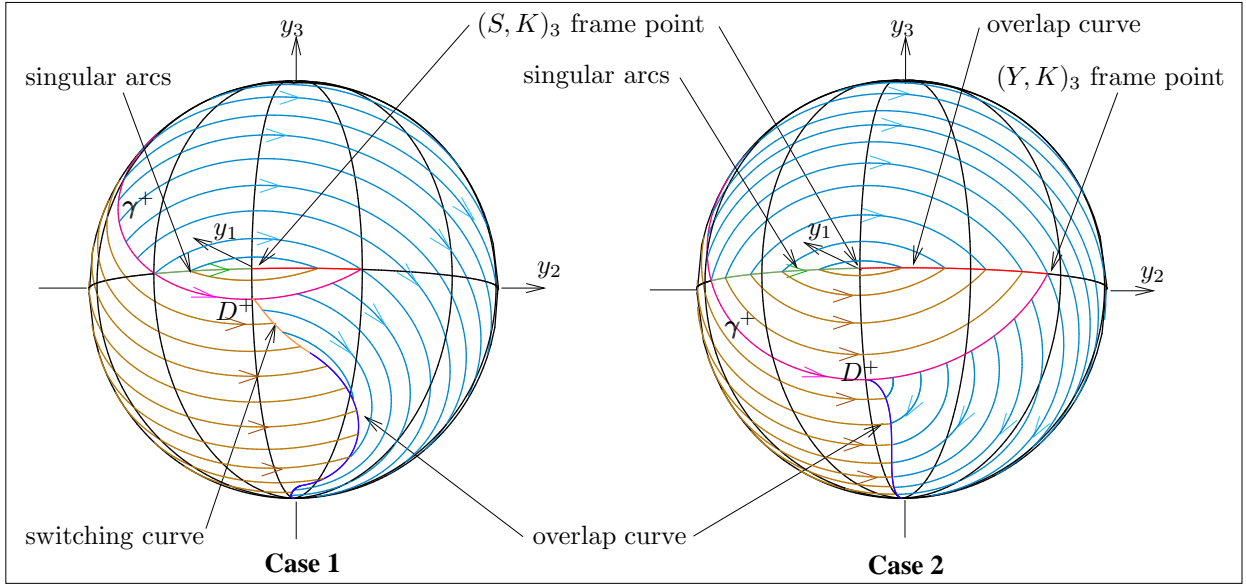}
\caption{Optimal synthesis for $\al=\pi/3$ and $\al$ 
slightly larger than $\pi/4$.}\llabel{f-sintesidoppia}. 
\end{center}
\end{figure} 

\subsubsection{The time optimal synthesis in $\ON^\pm$ and optimal trajectories reaching $P_S$
for $\al\geq\pi/4$}
From next proposition,  {\bf T11} of Theorem 
\ref{t-Evena} follows. 
More precisely Proposition \ref{propA} shows that 
in the case  $\al\geq\pi/4$, there are exactly four optimal trajectories 
steering $P_N$ to $P_S$, and it characterizes them. As a consequence, 
the south pole belongs to the cut locus.
\medskip
\bp
\llabel{propA}
Consider the control system \r{cs-p}--\r{cs-u}, and assume
$\al\geq\pi/4$. Then the optimal trajectories steering the north pole to the 
south pole are bang-bang with only one switching. More precisely they are 
the four trajectories corresponding to the four controls
{
\bqn
u^{(1)}=\left\{
\ba{l} 
u=1,~t\in[0,t_1]\\
u=-1,~t\in]t_1,T],\\
\ea
\right.~u^{(2)}=\left\{
\ba{l} 
1,~t\in[0,t_3]\\
-1,~t\in]t_3,T],\\
\ea
\right.~u^{(3)}=\left\{
\ba{l} 
-1,~t\in[0,t_1]\\
1,~t\in]t_1,T],\\
\ea
\right.~u^{(4)}=\left\{
\ba{l} 
-1,~t\in[0,t_3]\\
1,~t\in]t_3,T]\\
\ea
\right.\nn
\eqn
}
where $t_1$ and $t_3$ are defined in Definition \ref{d-Avena}, and $T=2\pi.$
\ep
\medskip
One can easily check that the switchings described in Proposition  
\ref{propA} occur on the equator ($y_3=0$).

The following proposition describes the optimal synthesis in $\ON^\pm$, in  a \neigh of the points 
$D^\pm$ and the bifurcation occurring at $\al=\arcsin(1/\sqrt[4]{2})$.
\bp
Let $\al\geq\pi/4$. In a \neigh of the point $D^+$  in  $\ON^+$, 
there exists a switching curve starting at $D^+$ of the form 
$e^{v(s)X_S^+}e^{sX_S^-}P_N$. If $\al>\pi/4$ this 
curve is tangent to the equator at  $D^+$.
Moreover 
if $\al<\arcsin(1/\sqrt[4]{2})$ (above called {\bf Case 1}) then the 
switching curve is   optimal near $D^+$, while if  
$\al\geq\arcsin(1/\sqrt[4]{2})$  (above called {\bf Case 2}) then the 
switching curve is not locally optimal near  $D^+$ and an overlap curve 
starts at the point $D^+$. A symmetric result holds in a \neigh of $D^-$ 
in  $\ON^-$.
\llabel{sw-over}
\ep

The region $\ON^+$ contains a cut locus that should be 
determined numerically.  
In {\bf Case 2}, numerical simulations show that the switching curve starting at $D^+$ is never 
optimal, i.e. every point of $\ON^+$ is reached by an optimal trajectory of the form $e^{t X_+} e^{s 
X_-}P_N$, with $s\in]0,t_1[$ or an 
optimal  trajectory of  the form  $e^{t X_-} e^{s
X_+}P_N$, with $s\in]\pi,t_3[$. 

\brem \llabel{r-nocal}
Notice however that, in {\bf Case 2}, given a point $\bar y\in\ON^+$, to find the time optimal trajectory 
reaching $\bar y$, it is not necessary to compute the cut locus. Indeed it is sufficient 
to compare the final 
times, corresponding to the two switching strategies given above,
and to chose the quickest one. The situation in $\ON^-$ is symmetric.
\erem

In  {\bf Case 1}, the situation is more complicated. 
The switching curve described by Proposition \ref{sw-over} has the expression
$C_1^{+}(s)= e^{X^{+}_Sv(s)}   e^{X^{-}_S s}  P_N$, $s\in]0,t_1[$
where the function $v(.)$ is given by the same formula of the $\al<\pi/4$ case, i.e. 
$v(s)=\pi+2 \arctan\left[{(\sin s)}/({\cos s+\cot^2 \alpha })\right].$ 
(To verify such formula it is enough to repeat the
computations done in \cite{y2}.) 
As described by  Proposition \ref{sw-over}, this switching curve  is optimal near $D^+$ and 
numerical simulations show 
that there exists $\bar s\in]0,t_1[$ such that
there is an optimal trajectory switching on 
 $C_1^{+}(s)$ if and only if  $s\in [0,\bar s[$,
and
an overlap 
curve connecting $C_1^{+}(\bar s)$  to the south pole appears.
The optimal synthesis for {\bf Case 1} and {\bf Case 2} 
 is depicted in Figure \ref{f-sintesidoppia}.


\subsection{Optimal trajectories reaching the south pole for $\al<\pi/4$}
\llabel{s-traj-polo}
In this section we characterize the time optimal trajectories reaching the south pole, in the case 
$\al<\pi/4$. This characterization is more complicated with respect to the 
case $\al\geq\pi/4$, due to the fact that the optimal trajectories have many switchings. 
The time optimal synthesis for $\al<\pi/4$ was already 
(partially) studied in \cite{y2} and it has been described in Section \ref{hist}.

From conditions {\bf i)}--{\bf iiii)}
in Section \ref{hist}, we know that every optimal trajectory starting at 
the north pole has the form $B_{s_i}{B_{v({s_i})}\cdots 
B_{v({s_i})}}B_{s_f}$\vspace{0.1cm}
where the function $v(s_i)$ is given by formula \r{v()}. (In the following 
we do not specify if the first bang corresponds to control $+1$ or $-1$, 
since, as a consequence of the symmetries of the problem, if $u(t)$ is 
an optimal control steering the north pole to the south pole, $-u(t)$ 
steers the north pole to the south pole as well.)  
It remains to identify one or more values of $s_i,s_f$ and the 
corresponding  number of switchings $n$ for this trajectory to reach the 
south pole.

Notice that $\bar{t}=\arccos(-\tan^2(\al))$ is the maximum of the function $v(.)$ 
on the interval $[0,\pi]$, $v(.)$ is increasing on $[0,\bar{t}]$ and decreasing
on $[\bar{t},\pi]$ and $v(0)=v(\pi)=\pi$.
Then, given $s\in[0,\pi]$ such that $s\neq\bar{t}$, there is a unique solution 
$s^\ast(s)\in [0,\pi]$, $s^\ast(s)\neq s$, to the equation $v(s^\ast)=v(s)$. The function 
$s^\ast(.)$ is extended to the whole interval $[0,\pi]$ setting $s^\ast(\bar{t})=\bar{t}$
(see Figure \ref{f-vfg} A). 
Thanks to the symmetries of the problem, we prove 
that if $\al<\pi/4$,  $s_f$ is equal either to $s_i$ or to $s^\ast(s_i)$.
This fact  is described by Lemma \ref{v=v} stated and  proved in Appendix~\ref{a-pifferone}.
\begin{figure}
\begin{center}
\input{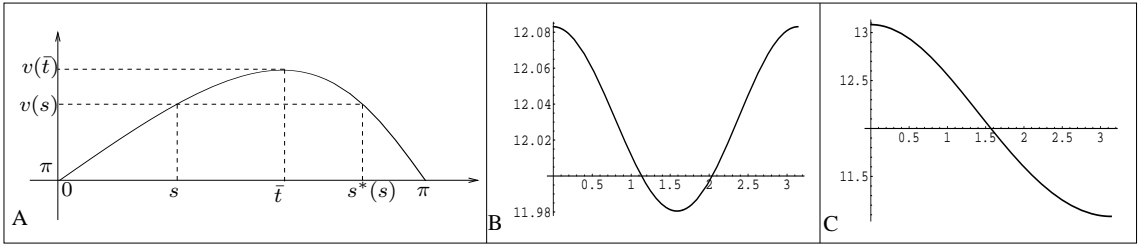}
\caption{Graph of $v(.)$ when $\al=\pi/6$ (figure A). Graph of the 
functions $\mathcal{F}$ and
$\mathcal{G}$ when $\al=0.13$ (figure B and C)}
\llabel{f-vfg}
\end{center}
\end{figure}

The following two propositions describe how to identify candidate 
triples $(s_i,s_f,n)$ for which the corresponding trajectory steers the 
north pole to the south pole in minimum time.
We say that a bang-bang trajectory, solution of the system \r{cs-p}--\r{cs-u},
is a \und{candidate optimal trajectory} if it is an extremal trajectory 
for problem {\bf (P)} reaching the south pole and it has a number $n$ of 
switchings satisfying $n\leq N_M$ (defined in Formula \r{eq-nmax}).
From Lemma \ref{v=v}, there are two kinds of candidate optimal trajectories: 
\bi
\i $s_f=s^\ast(s_i)$, called TYPE-1-candidate optimal trajectories
\i $s_f=s_i$ called TYPE-2-candidate optimal trajectories
\ei
\vspace{0.1cm}
Define the following functions, whose geometric meaning is clarified in
Appendix~\ref{a-pifferone2}:
\bqn
\th(s)\!=\!2 \arccos\!\left(\sin^2\!\left(\frac{v(s)}{2}\right)
\cos (2\al)-\cos^2\!\left(\frac{v(s)}{2}\right)\right)\!\!
\eqnl{thth}
\bqn
\beta(s)=2 \arccos (\sin(\al)\cos(\al) (1-\cos(s)))
\eqnl{bebe}
\bp \llabel{p-1}{\bf (TYPE-1-trajectories)}
{\sl Fixed $\al<\pi/4$, the equation for the 
couple $(s,n)\in [0,\pi]\times \N$:
\bqn
\mathcal{F}(s):=\frac{2\pi}{\th(s)}=n,
\eqnl{effe}
has either two or zero solutions. More precisely if $(s,n)$ is a 
solution to equation \r{effe}, then $(s^\ast(s),n)$ is the second one.  
The trajectories $B_{s}\underbrace{B_{v({s})}\cdots 
B_{v({s})}}_{n-1}B_{s^\ast(s)}$ and $B_{s^\ast(s)}\underbrace{B_{v({s})}\cdots
B_{v({s})}}_{n-1}B_{s}$  are the
TYPE-1-candidate 
optimal trajectories.} 
\ep
\bp \llabel{p-2} {\bf (TYPE-2-trajectories)}
{\sl Fixed $\al<\pi/4$, the equation for the couple 
$(s,n)\in [0,\pi]\times\N$:
\bqn
\mathcal{G}(s):=\frac{2\beta(s)}{\th(s)}+1=n,
\eqnl{gi}
has exactly two solutions. More precisely these solutions have the form 
$(s_1,n)$, $(s_2,n+1)$. The trajectories 
$B_{s_1}\underbrace{B_{v({s_1})}\cdots B_{v({s_1})}}_{n-1}B_{s_1}$ and  
$B_{s_2}\underbrace{B_{v({s_2})}\cdots B_{v({s_2})}}_{n}B_{s_2}$ 
are the TYPE-2-candidate optimal trajectories.}
\ep
In Figure \ref{f-vfg} B and C the graphs of the functions \r{effe} and 
\r{gi} 
are drawn for a particular value of $\al$, namely $\al=0.13$.
Propositions \ref{p-1} and \ref{p-2} select a set of (possibly coinciding) 
$4$ or $8$ candidate optimal trajectories (half of them starting with control $+1$ and the 
other half with control $-1$) corresponding to triples $(s_i,s_f,n)$. Such triples  
can be easily computed numerically solving equations \r{effe} and \r{gi}.
Then the optimal trajectories can be selected by comparing the times needed 
to reach the south pole for each of the candidate optimal trajectory. 
Notice that there 
are at least two optimal trajectories steering the north to the south 
pole (one starting with control $+1$ and the other with control $-1$).

If $\pi/(2\al)$ is an integer number $\bar{n}$, then TYPE-1 candidate 
optimal trajectories 
coincide with the TYPE-2  candidate optimal trajectories of the form  $B_\pi \underbrace{B_\pi 
...B_\pi}_{\bar{n}-2}B_\pi$. The remaining trajectories of TYPE-2 are of the form
$B_s \underbrace{B_{v(s)} ...B_{v(s)}}_{\bar{n}-1} B_s$ for some $s\in]0,\pi[$.
Otherwise if $\pi/(2\al)$ is not an integer number, define:
$$
m:=[\frac{\pi}{2\al}],~~~\mbox{and 
the normalized remainder}~~~\rr:=\frac{\pi}{2\al}-\left[\frac{\pi}{2\al}\right]\in[0,1[.
$$
where $[.]$ denotes the integer part. The following proposition determines
precisely the time optimal trajectories for particular values of the parameter 
$\rr$:
\bp
\llabel{alterna}
For $m$ large enough there exist $r_1(m)\leq r_2(m)\in]0,1[$ such that:
\bd
\i[A.] if $\rr\in]0,r_1(m)]$ then equation \r{effe} admits exactly two 
solutions  that are both optimal, while TYPE-2 candidate optimal 
trajectories are not.

\i[B.] if  $\rr\in]r_1(m),r_2(m)[$, then equation  \r{effe} admits two 
solutions, that are not optimal.

\i[C.] if $\rr\in]r_2(m),1[$ then equation \r{effe} does not admit any 
solution. Moreover $r_2(m)\to0$ for $m\to\infty$.

\ed
\ep
\ffoot{Si puo` dimostrare che per $\rr$ grande ci sono solo 2 traiettorie ottime, mentre per $\rr$
piccolo ce ne sono 4}

\brem
The function $r_2(m)$ can be determined explicitly (see Appendix 
\ref{s-alterna}), while for $r_1(m)$ we are just able to prove the 
existence, and we 
conjecture that it can be taken equal to $r_2(m)$. 
\erem

\brem
\llabel{r-conj}
An important consequence of Proposition~\ref{alterna} is that for $\al$ small, 
the number of optimal trajectories reaching the south pole is not fixed 
with respect to  $\al$. Indeed such number alternates as $\al\to0$, 
according to  
Proposition~\ref{alterna}:  in 
particular it is equal to $4$ if $\rr\in]0,r_1(m)]$ and it is equal to $2$ 
if
$\rr\in]r_2(m),1[\cup\{0\}$.  This is enough to conclude that also the 
qualitative 
shape of the optimal synthesis in a neighborhood of the south pole 
alternates giving a partial proof to the conjecture {\bf C2} of  Section~\ref{hist} (originally 
stated in \cite{y2}). 
In particular it is a proof of the first assertion (on the dependence of 
the synthesis on the remainder $r=2\al\rr$). Moreover notice 
that  the results of Proposition~\ref{alterna} perfectly fit with all 
the other statements of conjecture {\bf C2} with $r_2(m)$ playing the role 
of $\al_1/(2\al)$. 
One can apply the definition of locally equivalent syntheses given  in  \cite{libro} (see 
Definition 32, pag. 59), to make 
rigorous the  statement that  the qualitative
shape  of the optimal synthesis  changes with $\al$.

\erem

Using the previous analysis one can easily show the following result (of which we skip the 
proof):\ffoot{non sarebbe meglio scriverla con la parte intera?}
\bp
{\sl
If $N$ is the number of switchings of an optimal trajectory joining the 
north to the south pole, then
$$\frac{\pi}{2\al}-1\leq N<\frac{\pi}{2\al}+1.  $$}
\ep

Using these inequalities and the fact that, for $\al<\pi/6$, the function
$\ \displaystyle{2s+\left(\frac{\pi}{2\al}-1\right)v(s)}\ $ is increasing on 
$[0,\pi]$, one can give a rough estimate of the time 
needed to reach the south pole:\ffoot{la dimostriamo???}

\bp
{\sl The total time $T$ of an optimal trajectory joining the north to the 
south 
pole satisfies the inequalities:
$$
\frac{\pi^2}{2\al}-2\pi<T<\frac{\pi^2}{2\al}+\pi. 
$$}
\ep

\subsection{Comparison with results in the rotating wave approximation and with \cite{brokko}}
\llabel{s-comparison}
In this section we come back to the original value of $k$ i.e. 
$k=2E/\cos(\al)=2\sqrt{M^2+E^2}$, and 
we compare the time necessary to steer the state one to 
the state two  for our model and  the model \r{eq-hgcomplex}, described 
in Remark~\ref{r-rwa}, in which  
we control the magnetic field both along the $x$ and $y$ direction, or we consider a two-level molecule 
in the rotating wave approximation. We recall that $-E,E$ are the energy levels and 
$M$ is the bound on 
the control.  For our model, the time of transfer  $T$ satisfies:
\bi
\i for $\al\geq\pi/4$ (i.e. for $M\geq E$) then $T=2\pi/k=\pi/\sqrt{M^2+E^2}$;
\i for $\al<\pi/4$   (i.e. for $M<E$) then $T$ is estimated by 
$\displaystyle 
\frac1k\left(\frac{\pi^2}{2\al}-2\pi\right)<T<\frac1k
\left(\frac{\pi^2}{2\al}+\pi\right).
$
\ei
On the other hand, for the model \r{eq-hgcomplex},  the time of transfer is 
$T_{\C}=\pi/(2M)$  (cfr. Remark~\ref{r-rwa}). 
\begin{figure}
\begin{center}
\input{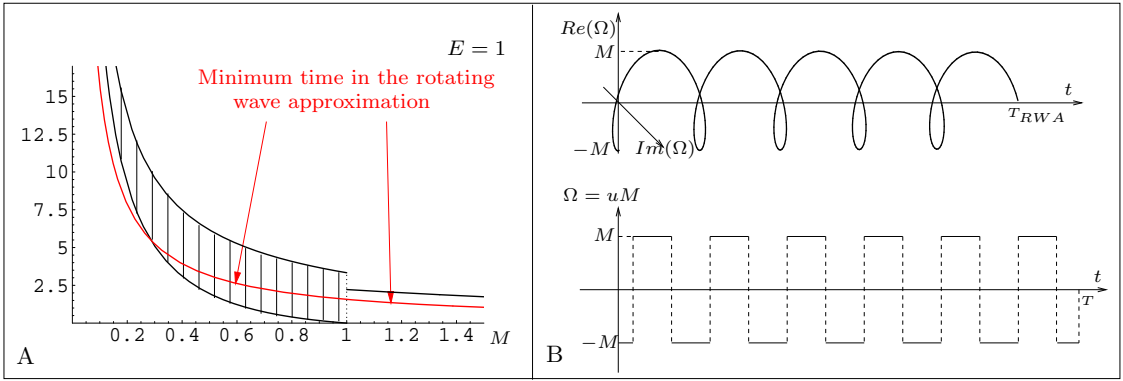}
\caption{~A. Estimate of the minimum time to reach the state two and 
comparison with the time needed with two controls or in the rotating wave approximation~~~ B. 
Comparison between the optimal strategy for our system and in the rotating wave approximation}
\llabel{f-innominata}
\end{center}
\end{figure}
Fixed $E=1$, in Figure \ref{f-innominata} A the times $T$ and 
$T_{\C}$ as function of $M$ are compared. Notice that although $T_{\C}$ is bigger than the lower 
estimate of $T$  in some interval, we always have $T_{\C}\leq T$. This is 
due to the fact that the admissible velocities of our model are a subset 
of the admissible velocities 
of the model \r{eq-hgcomplex}. 

Notice that, fixed $E=1$, for $M\to 0$ we have  $T\sim \pi^2/(4 M)=(\pi/2) T_{\C}$, while 
for $M\to \infty$, we have   $T\sim \pi/M= 2 T_{\C}$. 

\brem
For $M<<E$ (i.e. for $\al$ small) the difference between two switching times is $v(s)/k\sim \pi/(2E)$. 
It follows that a time optimal trajectory connecting the north to the south pole 
(in the interval between the first and the last bang)
is periodic with period $P\sim\pi/E$ i.e. with a frequency of the order 
of the resonance frequency $\om_R=2E$ (see Figure \ref{f-innominata} B).
On the other side if $M>E$ then the time optimal trajectory connecting the 
north with the south pole is the concatenation of two pulses. Notice that 
if $M>>E$, the time of transfer is of the order of $\pi/M$ and 
therefore tends to 
zero as $M\to\infty$. It is interesting to compare this result with a  
result of Khaneja, Brockett and Glaser, for a two level system, but 
with no bound on controls (see \cite{brokko}).
They estimate the infimum time to reach every point of the whole group $SU(2)$ 
in $\pi/E$. On the other side, in Appendix \ref{a-c} it is proved that the time needed to cover the whole 
sphere $\SB=SU(2)/S^1$ goes to $\pi/(4E)$ as $M$ goes to infinity (however this does
not contradict the fact that the state two can be reached in an arbitrary small
time, as we discussed above).


Notice that our optimal control has the same form of the control computed in 
\cite{brokko} i.e. a pulse (bang) followed by an evolution with the drift (singular) followed by 
a pulse (bang).

\erem




\appendix
\section{An overview on Optimal Synthesis on 2-D Manifolds}
\llabel{a-synt}
In this section we briefly recall the theory of optimal syntheses on 
2-D manifolds for system of the kind $\dot y = F(y) + u G(y)$, $|u|\leq 
1$, developed  by
Sussmann, Bressan, Piccoli and the first author in
\cite{ex-syn,quattro,due,sus2}
and recently rewritten in
\cite{libro}. This appendix is written to be as much self-consistent as 
possible.

For every coordinate chart on the manifold it is possible to introduce 
the following three functions:
\bqn
\Delta_A(y)&:=&Det(F(y),G(y))=F_1(y) G_2(y)-F_2(y)
G_1(y), \llabel{deltaA}\\
\Delta_B(y)&:=&Det(G(y),[F,G](y))=G_1(y)[F,G]_2(y)-G_2(y)[F,G]_1(y),
\llabel{deltaB}\\
f_S(y)&:=&-\Delta_B(y)/\Delta_A(y).\llabel{fs}
\eqn
The sets $\da,\db$ of zeroes of $\Delta_A,\Delta_B$ are respectively
the set of points where $F$ and $G$ are parallel, and the set of
points where $G$ is parallel to $[F,G]$.  These loci are fundamental
in the construction of the optimal synthesis.  In fact, assuming that
they are smooth embedded one dimensional submanifold of $M$ we have
the following:
\bi
\i in each connected region of $M\setminus(\da\cup \db)$, every extremal
trajectory  is bang-bang with at most one switching. Moreover, for
every switching of the extremal trajectory the value of the control passes
from $-1$ to $+1$  if $f_S>0$ and from $+1$ to $-1$
if $f_S<0$;

\i the support of singular trajectories (that are trajectories for which
the switching function identically vanishes, see Definition \ref{d-sw-f}
below) is always
contained in the set $\db$;

\i a trajectory not switching on the set of zeroes of $G$ is an abnormal
extremal (i.e. a trajectory for which the Hamiltonian given by the PMP
vanishing, see below) if and only if it switches on the locus $\da$.
\ei

Then the synthesis is built recursively on the number of switchings of
extremal trajectories, canceling at each step the non optimal
trajectories (see \cite{libro}, Chapter 1).\\
\brem
Notice that, although the functions $\Delta_A$ and  $\Delta_B$ depend on the coordinate chart,
the sets $\da$, $\db$ and the function $f_S$ do not, i.e. they are intrinsic objects of the
control equation $\dot y = F(y) + u G(y)$. 
\erem
\subsection{Basic Definitions and PMP on a $n$-dimensional Manifold} 
\llabel{a-pmp}
In this section we define our optimization problem, we state the Pontryagin Maximum Principle, 
and we give 
some basic definitions in the more general case of a $n$-dimensional manifold. 
We do this, since in Appendix~\ref{a-pifferone1} 
we stated some result for the original 
problem \r{eq-norm}, on $S^3\sim SU(2)$.\\\\
{\bf Problem (Q)} {\it
Consider the control system:
\bqn \dot y = F(y) + u G(y),~~~y\in M,~~~|u|\leq 1,
\eqnl{sys-gen}
where:
\bd
\i[(H0)] $M$ is a smooth $n$-dimensional manifold. The vector fields
$F(y)$ and $G(y)$ are
$\con^\infty$. 
\ed
We are interested in the problem of \und{reaching every point
of $M$ in minimum time} from a point $y_0\in M$.}

\bdeff
An admissible control $u(.)$ for the system \r{sys-gen} is a 
measurable function $u(.): [a,b]\to [-1,1]$,
while an admissible trajectory is a  Lipschitz functions 
$y(.):[a,b]\to M$ satisfying 
$\dot
y(t)=F(y(t))+u(t)G(y(t))$
a.e. for some  admissible control $u(.)$ 
\edeff
In the following we assume that the control system is 
\und{complete} i.e. for every measurable control
function $u(.):[a,b]\to [-1,1]$ and every initial state $\bar y$, there 
exists a trajectory $y(.)$ corresponding to $u(.)$, which is defined on the 
whole interval $[a,b]$ and satisfies $y(a)=\bar y$.
For us a solution to the problem {\bf (Q)} is an optimal synthesis. For a more elaborated 
definition of synthesis, see Section \ref{a-frame}.
\bdeff {\bf (Optimal Synthesis)} An optimal synthesis for the problem 
{\bf (Q)} is a
collection of time optimal trajectories
$\Gamma=\{y_{y_1}(.):[0,b_{y_1}]\mapsto M $, $y_1\in M :~~y_{y_1}(0)=y_0,
~y_{y_1}(b_{y_1})=y_1\}$.
\edeff
The key tool is the PMP (see \cite{agra-book,libro,jurd-book}).\\\\ 
\noi{\bf Theorem (Pontryagin Maximum Principle for the problem (Q))}
{\it Consider the  control system \r{sys-gen} subject to {\bf (H0)}.
Define for every $(y,\lam,u)\in T^\ast M\times [-1,1]$ the function
\bqn
\mathscr{H}(y,\lam,u)&:=&<\lam,F(y)>+u<\lam,G(y)>.
\eqnn
If the couple $(y(.),u(.)):[0,T]\to M\times [-1,1]$
is time optimal then there exist a \underline{never vanishing}
Lipschitz continuous \und{covector}
$\lam(.):t\in[0,T]\mapsto \lam(t)\in
T^\ast_{y(t)}M$ and a constant $\lam_0\leq 0$ such that for a.e. $t\in 
[0,T]$:
\begin{description}
\item[i)]
$\dot y(t)=\displaystyle{\frac{\partial \mathscr{H}
}{\partial \lam}(y(t),\lam(t),u(t))}$,
\item[ii)] $\dot \lam(t)=-\displaystyle{\frac{\partial \mathscr{H}
}{\partial y}(y(t),\lam(t),u(t))=-\lam(t)\big(\nabla
F(y(t))+u(t)\nabla G(y(t))\big)}$,
\item[iii)] $\mathscr{H} 
(y(t),\lam(t),u(t))=\mathscr{H}_M(y(t),\lam(t))$\mbox{ where }
$\mathscr{H}_M(y,\lam):=\max\{\mathscr{H}(y,\lam,u): 
u\in [-1,1]\},$
\item[iiii)] $\mathscr{H}_M(y(t),\lam(t))+\lam_0=0$.
\ed
}
\brem
\llabel{r-postPMP}
The PMP is just a necessary condition for optimality. A trajectory $y(.)$
(resp.  a couple $(y(.),\lam(.))$)
satisfying the conditions given by the PMP is said to be an \und{extremal} (resp.  an \und{extremal
pair}).
An extremal  corresponding to $\lam_0=0$ is said to be an \und{abnormal extremal}, otherwise we call it a 
\und{normal extremal}.
\erem
We are now interested in determining the extremal trajectories satisfying 
the conditions given by the PMP. A key role is played by
the following:
\bdeff {\bf (switching function)}
Let $(y(.),\lam(.))$ be an extremal pair. The corresponding
switching function is defined as $\phi(t):=<\lam(t),G(y(t))>$.
\llabel{d-sw-f}
\edeff
Notice that  $\phi(.)$ is continuously differentiable (indeed $\dot
\phi(t)=<\lam(t),[F,G](y(t))>$, that is continuous). 
\bdeff {\bf (bang, singular)} 
\llabel{d-BS-2}
Let $\g$, defined in $[a,b]$, be an extremal trajectory and 
$u(.):[a,b]\to[-1,1]$ the 
corresponding control. We say that $u(.)$  is a 
\und{bang} control if 
$u(t)=+1$ 
a.e. in  $[a,b]$ or  $u(t)=-1$ a.e.  in  $[a,b]$. We say that  
$u(.)$ is  
\und{singular}  if the corresponding switching function 
$\phi(t)=0$ in  $[a,b]$. A finite concatenation 
of bang controls is 
called a \und{bang-bang} control. A \und{switching time} of $u(.)$ is a 
time 
$\bar t\in[a,b]$ such that, for every $\vep>0$, $u$ is not bang or 
singular on 
$(\bar t-\vep,\bar t+\vep)\cap 
[a,b]$.
An extremal trajectory of the control system
\r{sys-gen} is said a bang extremal, singular extremal, bang-bang extremal 
respectively, if it
corresponds
to a bang control, singular control, bang-bang control
respectively. If $\bar t$ is a switching time, the corresponding point on 
the trajectory $y(\bar t)$ is called a \und{switching point}.
\edeff
The switching function is important because it  determines
where the controls may
switch. In fact, using
the PMP, one easily gets:
\bp A necessary condition for a time
$t$ to be a switching is that $\phi(t)=0$.
Therefore, on any
interval where
$\phi$ has no zeroes (respectively finitely many zeroes), the
corresponding control is bang (respectively bang-bang). In particular,
$\phi>0$ (resp $\phi<0$) on $[a,b]$ implies $u=1$ (resp.  $u=-1$)
a.e. on  $[a,b]$. On the other hand,
if $\phi$ has a zero at $t$ and $\dot \phi(t)$ is different
from zero, then $t$ is an isolated switching.
\llabel{p-bangs}
\ep
\subsection{More on singular extremals and predicting switchings for 
2-D systems}
\llabel{a-sing}
Now we come back to the case in which $M$ is two dimensional. In this 
Section we compute the control corresponding to singular extremals and 
we would like to predict which kind of switchings can happen, using
properties of the vector fields $F$ and $G$. 
The following two lemmas illustrate the role of the functions 
$\da$, $\db$ in relation with  singular 
 and abnormal extremals. The proofs can be found in
\cite{ex-syn,libro,due}.
\bl
\llabel{l-sing}
Let $y(.)$ be an extremal trajectory that is singular in
$[a,b]\subset Dom(y(.))$. Then $y(.)|_{[a,b]}$ corresponds to the so 
called
\underline{singular
control} $\varphi(y(t))$, where:
\bqn
\llabel{feedbk-varphi}
\varphi(y)=-\frac{\nabla \Delta_B(y)\cdot F(y)}{\nabla\Delta_B(y)\cdot
G(y)},
\eqn
with $\Delta_A$ and $\Delta_B$ defined in Eqs. \r{deltaA} and
\r{deltaB}. Moreover, on $Supp(y(.))$, $\varphi(y)$ is always well-defined
and its absolute value is
less than or equal to one.
Finally $Supp(y(.)|_{[a,b]})\subset \db$.
\el
\bl
\llabel{l-ab}
Let $y(.)$ be a bang-bang extremal for the
control problem (\ref{sys-gen}), $t_0\in Dom(y(.))$ be a time such that
$\phi(t_0)=0$ and $G(y(t_0))\neq0$.
Then, the following conditions are equivalent:
{\bf i)} $y(.)$ is an abnormal extremal;
{\bf ii)} $y(t_0)\in\da$;
{\bf iii)} $y(t)\in\da$, for every time $t\in Dom(y(.))$ such that
$\phi(t)=0$.
\el

The following lemma describes what happens when $\Delta_A$ and
$\Delta_B$ are different from zero.
\ffoot{Mi sembra proprio che sia scambiato -1 con +1: vedi teorema 11
pag 44 del libro}
\bl
\llabel{l-sw}
Let   $\Omega\subset M$ be an open set such that
$\Omega\cap (\da\cup \db)=\emptyset$.
Then all connected components of $Supp(y(.))\cap \Omega$, where $y(.)$ is 
an
extremal trajectory of (\ref{sys-gen}), are
bang-bang with at most one switching. Moreover, if $f_S>0$ throughout
$\Omega$, then $y(.)|_{\Omega}$ is associated to a \cc equal to $+1$ or 
$-1$
or has a  switching from $-1$ to $+1$. If $f_S<0$ throughout $\Omega$,
then $y(.)|_{\Omega}$ is associated to a \cc equal to $+1$ or $-1$ or has 
a switching from $+1$ to $-1$.
\el

\subsection{Frame Curves and Frame Points}
\llabel{a-frame}
For the problem {\bf (Q)}, under generic conditions on the vector fields 
$F$ and $G$, one can make the complete classification of synthesis 
singularities, stable synthesis, singularities of the minimum time wave 
fronts. 

In the following, for sake of completeness, we recall the main
results on existence of an optimal synthesis and on classification of synthesis singularities
obtained in \cite{uno,due}  (see  also \cite{libro}). 
In \cite{uno}, it  
was proved that the control system \r{sys-gen}, under generic conditions 
on $F$ and $G$ (with the additional assumption $F(y_0)=0$)
admits a time optimal \underline{regular synthesis} in finite time $T$,
starting
from $y_0$.
By generic conditions, we mean conditions verified on an open and
dense subset of the  set of $\con^\infty$ vector fields endowed with
the $\con^3$ topology (see \cite{libro}, formula 2.6 pp. 39).
To define what we mean by regular synthesis, we first need to introduce 
the concept of \und{reachable set} and of \und{stratification} of the 
reachable set.
We call reachable set in time $T>0$, the set: 
\bqn
{\cal R}(T)&:=&\{y\in M :\ \exists ~b_y\in[0,T]\mbox{ and a trajectory
}\nn\\&&\ga_y:[0,b_y]\to M\mbox{ of (\ref{sys-gen}) such that
}\ga_y(0)=y_0,~
\ga_y(b_y)=y\}.
\eqnn
Then we need the definition of stratification of  ${\cal R}(T)$, 
that roughly
speaking is a partition in manifolds of different
dimensions.
\bdeff {\bf (stratification)}
A stratification of $\reachT$, $T>0$, is a finite collection
$\{M_i\}$ of connected embedded $\con^1$ submanifolds of $M$,
called strata, such that  ${\cal R}(T)=\cup_k M_k$, and
the following holds. If $M_j\cap Clos(M_k)\not=\emptyset$ with $j\not= k$
then $M_j\subset Clos(M_k)$ and $dim(M_j)<dim(M_k)$.
\edeff
Then a \underline{time optimal regular synthesis} is defined by: {\bf i)}
a family of time optimal
trajectories
$\Gamma=\{\ga_y:[0,b_y]\to M $, $y\in{\cal
R}(T):~~\g_y(0)=y_0, ~\g_y(b_y)=y\}$
such that if $\g_y\in\Gamma$ and $\bar y=\gamma_{y(t)}$ for some $t\in
[0,b_y]$, then $\gamma_{\bar y}=\gamma_y|_{[0,t]}$; {\bf ii)} a
stratification of ${\cal R}(T)$ 
such that the optimal trajectories of $\Gamma$ can be
obtained from a feedback $u(y)$ satisfying:
\bi
\item on  strata of
dimension 2, $u(y)=\pm 1$,
\item  on  strata of dimension 1, called \underline{frame curves}
(FC for short), $u(y)=\pm 1$ or
$u(y)=\varphi(y)$, where $\varphi(y)$ is defined by 
(\ref{feedbk-varphi}).\ei
The strata of dimension 0
are called \underline{frame points} (FP).
Every  FP  is an intersection of two FCs.
\ffoot{c'era: An FP $y$, which is the
intersection
 of  two frame curves $F_1$ and $F_2$ is called an
$(F_1,F_2)$ Frame Point.}
 In \cite{due} (see also \cite{libro}), it is
provided a complete classification of all types of FPs and FCs, under
generic conditions. All the
possible FCs are:
\bi
\i FCs of kind $Y$ (resp. $X$), corresponding to
subsets
of the trajectories $\g^+$ (resp. $\g^-$) defined as the trajectory
starting at $y_0$ with \cc $+1$ (resp. \cc $-1$);
\i FCs of kind $C$, called {\sl switching curves}, i.e. curves made of
switching points;
\i FCs of kind $S$, i.e. singular extremals;
\i FCs of kind $K$,  called overlaps and reached optimally by two
trajectories coming from different directions;
\i FCs  which are  arcs of optimal trajectories starting
at FPs. These trajectories ``transport''  special information.
\ei
The FCs of kind $Y,C,S,K$ are depicted in Fig.~\ref{fig-FCs}.
There are eighteen topological equivalence classes
of FPs. A detailed description can be found in \cite{morse,libro,due}.
\ppotR{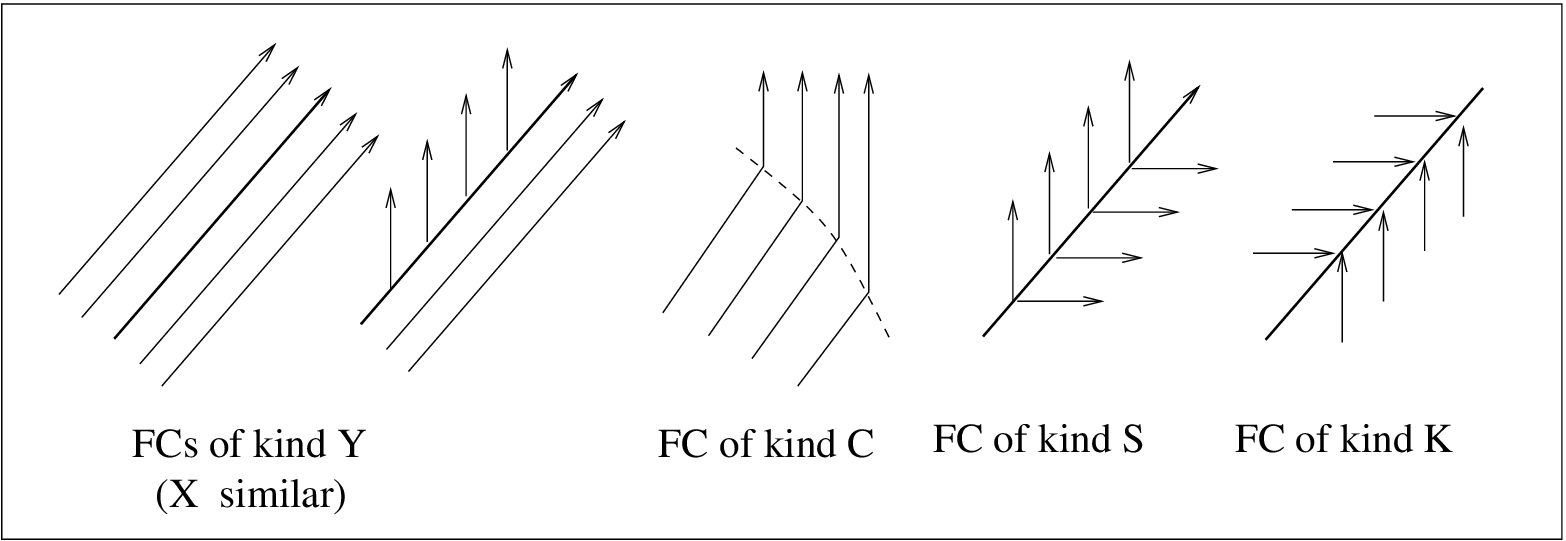}{Frame Curves of the optimal synthesis, under generic 
conditions}{9}

\section{Proof of the Main Results}
\llabel{a-pifferone}

In this section we give the  proof of our main
results. 
We start with a lemma, stating 
a property of optimal trajectories, that is  a consequence of the 
symmetries of the 
problem.  It is used to identify the time optimal 
trajectories steering the north to the south pole both for $\al\geq\pi/4$ and $\al<\pi/4$.
\bl
\llabel{v=v}
Let $\al\in]0,\pi/2[$.  Every optimal bang-bang trajectory, 
connecting the north to the south pole, with more than one switching is such
that $v(s_i)=v(s_f)$ where $s_i$ is the first switching time, $s_f$ is the 
time needed to steer the last switching point to the south pole and $v(s_i)$ is the time between 
two consecutive switchings.
\el
{\bf Proof of Lemma \ref{v=v}.} Consider the problem  of connecting $P_S$ with $P_N$
in minimum time for the system $\dot{z}=F'_S(z)+uG'_S(z)$
where $z\in S^2 $ and $F'_S(z)=-F_S(z)$, $G'_S(z)=-G_S(z)$.
The trajectories of this system  coincide with those of the 
system \r{cs-p}--\r{cs-u}, but the velocity is reversed. Therefore the optimal 
trajectories for the new problem coincide with the optimal ones for the
system \r{cs-p}--\r{cs-u} connecting $P_N$ to $P_S$, and the 
time between two switchings is the same. Since performing the change of coordinates 
$(z_1,z_2,z_3)\rightarrow (y_1,y_2,y_3)=(-z_1,z_2,-z_3)$, the new problem becomes exactly the 
original problem, we deduce that, if we have more than one switching, 
it must be $v(s_{i})=v(s_{f})$. \quadp
\subsection{Time Optimal Synthesis for the two Level Quantum System for $\al\geq\pi/4$}
\llabel{a-pifferone1}
In this section, we apply the theory of optimal syntheses on
2-D manifolds recalled in Appendix \ref{a-synt}, to the system 
\r{cs-p}--\r{cs-u}. Our aim is to describe the time optimal synthesis for $\al\geq\pi/4$, i.e. to
prove Theorem \ref{t-Evena} and Propositions \ref{propA} and  
\ref{sw-over}.
First we state some general results, holding for $\al\in]0,\pi/2$[, regarding time optimal trajectories 
of the system \r{eq-norm}, on $S^3\sim SU(2)$, analogous to those obtained in \cite{y2} for $SO(3)$
(in particular the proofs can be repeated using the same arguments).
\subsubsection{General results on $S^3$}
In this section  $\al\in]0,\pi/2$[.
The first proposition states that singular extremals, defined as extremals 
for which the switching 
function  vanishes (see Definitions \ref{d-sw-f} and \ref{d-BS-2}) correspond to zero control. This 
fact is very specific for 
our problem.

\bp
For the normalized  minimum time problem on $S^3$ \r{eq-norm}, singular extremals
are integral curves of the drift, i.e. they must correspond to a 
control almost 
everywhere vanishing. 
\ep
Since for a fixed $u\in [-1,1]$ every trajectory of 
\r{eq-norm} is periodic with period $\frac{4\textstyle{\pi}}
{\sqrt{u^2\sin^2\al+\cos^2\al}}$ 
\ffoot{forse bisognerebbe sottolineare meglio il fatto che fisicamente sono 
equivalentigli stati $e^{i\phi}\psi$ in intro} we have that:
\bp
Given an extremal trajectory $\ga$ of type $B_t$ (resp $S_t$), then 
$t<4\pi$ (resp. $t<\frac{4\pi}{\cos\al}$).
\label{max}
\ep
The following proposition describes the switching behavior of 
abnormal and bang-bang normal extremals (see Section \ref{a-pmp} for the definition).
\bp
Let $\ga$ be an abnormal extremal of \r{eq-norm}. Then it is  bang-bang 
and the time duration between two
consecutive switchings is always equal to $\pi$. 
In other words, $\ga$ is of kind $B_sB_{\pi}...B_{\pi}B_t$ 
with $s,t\leq\pi$.\\
On the other hand, if $\gamma$ is a bang-bang normal extremal, then 
the time duration ${\cal T}$ along an
interior bang arc is the same for all interior bang arcs and verifies
$\pi<{\cal T}< 2\pi$ (i.e. $\ga$ is of kind $B_sB_{{\cal T}}...B_{{\cal T}}B_t$
with $s,t\leq {\cal T}$).
\ep 
For the optimal trajectories containing a singular arc we have the following:
\bp\label{sing}
Let $\gamma$ be a time optimal trajectory containing a singular arc.
Then $\gamma$ is of the type $B_tS_sB_{t'}$, with 
$s\leq\frac{2\pi}{\cos\al}$
if $t>0$ or $t'>0$ and $s<\frac{4\pi}{\cos\al}$ otherwise.
\ep
These results on $S^3\sim SU(2)$ are useful to determine the optimal synthesis on $\SB$, since every 
optimal trajectory on $\SB$ is the 
projection of an optimal trajectory on $S^3$. This is a simple consequence of the fact that $\SB$ is an 
homogeneous space of $SU(2)$:
\bp\label{project}
A time optimal trajectory $\ga$ for the system \r{cs-p}--\r{cs-u} on $\SB$ starting at $P_N$ is the 
projection of a time optimal trajectory of
\r{eq-norm} starting from a point satisfying $|\psi_1|^2=1$ (recall that 
$\psi=(\psi_1,\psi_2)^T\in S^3\subset\C^2$).
\ep
\brem
Notice that, since two opposite points on $S^3$ project on the same point
on $\SB$, it is easy to see from 
Proposition \ref{max}, that the projection on $\SB$ of an optimal trajectory 
of \r{eq-norm} of type $B_t$ (resp $S_t$), must be such that  
$t<2\pi$ (resp. $t<\frac{2\pi}{\cos\al}$). 
More precisely,  for a fixed $u\in [-1,1]$ every trajectory of
\r{cs-p}--\r{cs-u} is periodic with period $\frac{2\textstyle{\pi}}
{\sqrt{u^2\sin^2\al+\cos^2\al}}$ (the period divides by two after projection).
\erem

\subsubsection{Construction of the Synthesis on $\SB$}
In this section we assume $\al\geq\pi/4$. Following Appendix \ref{a-synt} we first need to determine the 
sets $\da$, $\db$, and the
function $f_S$.  Checking where $F_S$ is parallel to $G_S$ and where $G_S$ is parallel to $[F_S,G_S]$, 
one gets $\da=\{y\in\SB:y_2=0\}$ and $\db=\{y\in\SB:y_3=0\}$. 
To find the function $f_S$ we can choose for instance the coordinate chart defined on each hemisphere 
by the projection on the plain $\{(y_1,y_2)\in \R^2\}$, obtaining $f_S=(\sin\al) y_3/y_2$. 
Then Lemma \ref{l-sw} says that, every optimal trajectory belonging to one of the regions
$\{y\in\SB:y_3>0,y_2>0\}$, $\{y\in\SB:y_3<0,y_2<0\}$ is bang-bang with  at most one switching. 
Moreover only the switching from control  $-1$ to control $+1$ is allowed.
On the contrary, on the regions  
$\{y\in\SB:y_3>0,y_2<0\}$, $\{y\in\SB:y_3<0,y_2>0\}$, the control can switch only from 
$+1$ to $-1$. Moreover, thanks to Lemma~\ref{l-sing}, every singular 
extremal must lie on the equator. 
The following lemma characterizes the structure of the bang-bang extremals for the problem {\bf (P)}.
\bl
Recall that $t_1=\pi-\arccos(\cot^2\al)$ and 
$t_3=\pi+\arccos(\cot^2\al)$ and consider a bang-bang extremal for the problem {\bf (P)}. 
Then 
it is of the form $B_sB_{v(s)}B_{v(s)}\ldots$ with 
$s\in[0,t_1]\cup [\pi,t_3]$, where, on the set  $\,[0,t_1[\cup 
[\pi,t_3[\,$, 
$v(.)$ is defined as follows:
$$
v(s) := \pi+2 \arctan\left(\frac{\sin s }
{\cos s+\cot^2 \alpha }\right).
$$
If $\al=\pi/4$ then $t_1=t_3=\pi$ and $v(\pi):=\pi$, while if $\al>\pi/4$ we set
$v(t_1):=v(t_3):=2\pi$.
\llabel{bang-sw}
\el 
Notice that the function $v(.)$ has the same expression \r{v()} obtained in
the case $\al<\pi/4$ (excepted at the points   $t_1$ and $t_3$). However its interval of 
definition is different.\\\\
{\bf Proof of Lemma \ref{bang-sw}.}
As shown above, the meridian $\da$ and the equator $\db$ divide the sphere in
four parts and in each of them the sign of the function $f_S$ is constant and
changes when passing through $\da$ or $\db$.
In particular, following $\ga^+$ or $\ga^-$ (cfr Remark \ref{r-4}) in the case in which 
$\al>\pi/4$ 
this happens at the times $t_1$ (where the equator is crossed), at time  $\pi\,$  (where $\da$ 
is crossed) and
at time $\,t_3$ (again is the equator to be crossed). Applying 
Lemma~\ref{l-sw}, we obtain that for an extremal trajectory the first 
switching may occur only on the intervals $[0,t_1]\,$ and 
$\,[\pi,t_3]$.
Exactly as in \cite{y2}, one shows that the extremal must have the form
$B_sB_{v(s)}B_{v(s)}\ldots$ with $s\in[0,t_1]
\cup [\pi,t_3]$. 
The case $\al=\pi/4$ is similar.
\quadp
\brem One can also show that  every trajectory starting from $P_N$, of the form 
$B_sB_{v(s)}B_{v(s)}\ldots$ with $s\in[0,t_1]
\cup [\pi,t_3]$ is extremal i.e., for every $s$ in such set, there
exists an initial value of the covector $\lam$ such that the switching function 
$\phi(.)$ vanishes for the first time at time $s$. 
\erem
Unlike the case in which $\al<\pi/4$, in the case $\al>\pi/4$ it is possible to establish
the presence of optimal trajectories containing a  singular arc, whose switching strategy is described by 
the following proposition, illustrated in Figure \ref{f-sintesi-sing} A.
\bp
Let $\al\geq\pi/4$. A trajectory $\ga$ of \r{cs-p}--\r{cs-u} starting with control $u=1$ and containing a 
singular 
arc is a  solution of {\bf (P)} if and only if it is of the form $B_tS_sB_{t'}$ and satisfies 
the following conditions:
\bi
\i $t=t_1=\pi-\arccos (\cot^2\al)$ i.e. $\ga$ coincides with $\ga^+$ until it reaches the 
equator.
\i $s\leq\arccos (\cot\al)/\cos\al$ i.e. the singular arc is optimal until it reaches the point 
$O^+=(1,0,0)^T$.
\i If $s=\arccos (\cot\al)/\cos\al$, then the trajectory is of type $B_tS_s$ (i.e. the 
time duration of the last bang arc reduces to zero). If $s<\arccos (\cot\al)/\cos\al$, then 
$\g$ is optimal until the last bang arc reaches the equator (i.e. it does not exist $\bar 
t\in]0,t'[$ such that $\g(t+s+\bar t)$ is contained in the equator).
\ei
An analogous result holds for trajectories starting with control $-1$.
\llabel{sing-synt}
\ep
\brem
Notice that in the case $\al=\pi/4$, Proposition \ref{sing-synt} provides a singular trajectory 
degenerated to a point. In other words for $\al=\pi/4$ there are no singular trajectories that are 
optimal.
\erem

\brem
Notice that the previous result completely characterizes the optimal synthesis in some
neighborhoods of the points $O^\pm=(\pm1,0,0)^T$, namely $\Omega_2^\pm\cup\Omega_3^\pm$, and 
moreover it 
determines the presence of two symmetric overlap curves contained inside the equator. 
The synthesis around the point $O^+$ is represented in Figure \ref{f-sintesi-sing} A. 
\erem

\ppotR{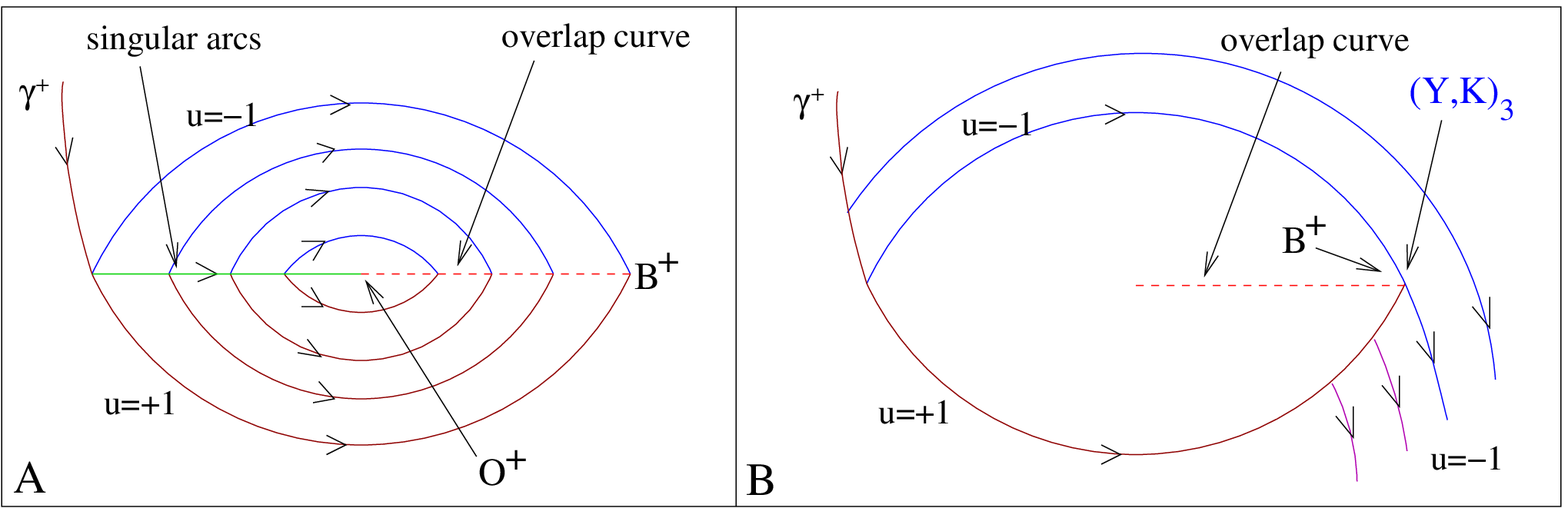}{The region covered by optimal trajectories with 
singular arcs and the $(Y,K)_3$ frame point}{12}
{\bf Proof of Proposition \ref{sing-synt}.}
Consider a trajectory, that is a solution of {\bf (P)} starting with $u=+1$ 
and containing a singular arc.
Using  Propositions~\ref{sing} and \ref{project} this trajectory must be of 
the form $B_tS_sB_{t'}$ and, since the singular arc is contained inside
the equator, we have $t=t_1$ (the case 
$t=t_3$ can be easily excluded).
\ffoot{per esempio se questa fosse ottima sarebbe ottima anche la traiettoria 
che swiccia da +1 a -1 al tempo $t_1$ a cui poi segue un 
tratto singolare cosa che pero' e' esclusa dalla Proposizione~\ref{sing}}
Consider a singular arc containing in its interior the point $O^+$.
This arc contains two points of the form $(y_1^0,-y_2^0,0)^T$ and 
$(y_1^0,y_2^0,0)^T$, with both $y_1^0$, $y_2^0$ positive, that can be 
connected by a bang arc. Using classical comparison theorems for second 
order ODEs, one can easily compare the time needed to follow
such trajectory with the time needed to steer the two points along the
singular arc finding that the bang arc is quicker than the singular arc.
Therefore a singular arc containing $O^+$ cannot be optimal.
By symmetry, the extremal trajectories that have the same singular arc, but 
the last bang arc 
corresponding to opposite control, must meet on a point of the equator. 
Therefore the arc of the equator which is comprised between the point $O^+$ 
(resp  $O^-$) and the second intersection point with $\ga^+$ (resp.  $\ga^-$) 
is an overlap curve. It remains now to verify that the trajectories described 
above are optimal (until the last bang arc reaches the equator). 
This is a straightforward consequence of the fact that the quickest 
bang-bang trajectories that enter the region spanned by such trajectories 
(i.e. the closure of the regions $\Omega^\pm_2\cup\Omega^\pm_3$) are not 
extremal because of Lemma~\ref{l-sw}. \quadp
\brem
\llabel{r-trivial-susman}
Notice the trivial fact that, if a trajectory $\g$ defined on the interval 
$[a,b]$ is optimal between $\g(a)$ and $\g(b)$, then the restriction of 
$\g$ in $[c,d]$, $c,d\in[a,b]$, $c<d$, is optimal between  $\g(c)$ and 
$\g(d)$. 
\erem

Using Remark \ref{r-trivial-susman}, we have that 
Proposition \ref{sing-synt} characterizes 
completely the time 
optimal synthesis on $\overline{P_NA^\pm}$ and in the closure of 
$\Omega_2^\pm\cup\Omega_3^\pm$, i.e. it proves 
items 
{\bf T1}--{\bf T6},
{\bf T9} and {\bf T10}, of Theorem \ref{t-Evena}.

\brem
From Lemma~\ref{bang-sw} we obtain that 
there are four families of bang-bang trajectories. In particular the families starting with control $+1$ 
and switching respectively in  $[0,t_1]$ and  $[\pi,t_3]$ join
at the point $B^+$, generating an amazing $(Y,K)_3$ frame point, in the framework of the classification 
of \cite{libro}. See Figure \ref{f-sintesi-sing} B.
\erem
Next we give the proof of Proposition  \ref{propA}, from which it follows {\bf T11} of Theorem \ref{t-Evena}, 
and,  using again Remark \ref{r-trivial-susman},  
also {\bf T7}.\\\\	
{\bf Proof of Proposition \ref{propA}}
By Proposition \ref{sing-synt}, there are no optimal trajectories containing a singular arc joining  
$P_N$ with $P_S$. One can easily see that  the only possible trajectories steering $P_N$ to $P_S$
with only one switching are those described in the statement of the proposition, that 
we have to compare with trajectories having more than one switching. 
Trajectories having two switchings with the first or the last bang longer than $\pi$ and 
trajectories with more than two switchings are excluded since from Lemma \ref{bang-sw} their total 
time is larger than $2\pi$. 
Trajectories  having two switchings and  length of the first arc $s_i$ and the length of the last arc 
$s_f$ satisfying  $s_i,s_f<\pi$ are excluded since by Lemma  \ref{v=v} they must  satisfy $s_i=s_f$.
For these trajectories the total time can be easily computed and it is $\,2\pi+2\arcsin
\left(\frac{1}{2\sin(\al)}\right)>2\pi$.
\quadp\\\\
Item  {\bf T8} is proved by the following: 
\bp
\llabel{p-porcamadosca}
If $\bar y\in\Omega_1^+\cup(\overline{DP_S^-}\setminus P_S)$, then  
$\GY$  is made by a unique trajectory of the form 
$B_tB_{t'}$, with  $0\leq t<t_1$ and  the first bang corresponding to 
control $+1$. A similar result holds if  $\bar y\in\Omega_1^-\cup(\overline{DP_S^+}\setminus 
P_S)$.
As a consequence there is not a cut locus in the region  $\Omega_1^+\cup\Omega_1^-$.  On the 
other hand $\ON^+\cup\ON^-$ contains a cut 
locus. 
\ep
{\bf Proof of Proposition \ref{p-porcamadosca}} Define the following three families of extremal trajectories:
\bqn
\g^A_s(t)&:=&e^{tX_S^+}e^{s X_S^-}P_N,\mbox{~~with~~}s\in]0,t_1[\mbox{~~and~~}t\leq v(s),\nn\\ 
\g^B_s(t)&:=&e^{tX_S^-}e^{s X_S^+}P_N,\mbox{~~with~~}s\in[\pi,t_3[\mbox{~~and~~}t\leq v(s),\nn\\ 
\g^C_s(t)&:=&e^{tX_S^-}e^{v(s)X_S^+}e^{s X_S^-}P_N,\mbox{~~with~~}s\in]0,t_1[\mbox{~~and~~}t\leq v(s).\nn 
\eqn
First notice that from Proposition \ref{propA}, there are no optimal trajectories of kind $\g^A_s$
reaching the arc $\overline{BP_S^+}$. 
Now for every point $x\in\overline{DP_S^+}$ the following happens: {\bf i)} there exist $s_A,~t_A$ such that 
$x=\g^A_{s_A}(t_A)$, and they are 
unique; {\bf ii)} if there exist $s_B,~t_B$ (resp.  $s_C,~t_C$)  such that
$x=\g^B_{s_B}(t_B)$,  (resp. $x=\g^C_{s_C}(t_C)$), then they are 
unique.  By direct computation, one can compare the times the three trajectories need to 
reach $x$, i.e. $s_A+t_A,s_B+t_B,s_C+v(s_C)+t_C$, finding that the optimal trajectory is of kind $\g^A$ (these computations are long, not very instructive, and we omit them). 
From this fact the first part of the claim immediately follows.
Moreover it implies that there is not a cut locus in $\Omega_1^+$, 
since the only trajectories entering 
such region are those of the form $\g^A$.
The existence of a cut locus in $\ON^+$ is evident, since no optimal trajectory belonging to the families 
$\g^A$,
$\g^B$,
$\g^C$ leaves $\ON^+$.
The reasoning in  $\Omega_1^-$ and in $\ON^-$ is similar. \quadp\\\\
{\bf End of the proof of Theorem \ref{t-Evena}}\\
To conclude the proof of Theorem  \ref{t-Evena}, it remains to prove 
{\bf T12}. Consider by contradiction an optimal bang-bang 
trajectory $\g$ defined in $[0,t_\g]$ 
steering $P_N$ to a point of  
$\ON^+$, with at least three switchings. 
Define $\bar{t}=\max \{t\in[0,t_\g]: \g(t)\notin \ON^+\}$. Then, by Remark~\ref{r-trivial-susman}, 
$\g|_{[0,\bar{t}]}$ must be optimal between $P_N$ and $\g(\bar{t})$. Then, from the results proved above, 
we deduce that $\g|_{[0,\bar{t}]}$ can have at most one switching. Therefore $\g$ switches at least two 
times in $\ON^+$ and the arc between them must be completely contained in $\ON^+$ and this leads to a 
contradiction since the sign of $f_S$ is constant in $\ON^+$ (see 
Lemma~\ref{l-sw}). \quadp\\\\
Before proving Proposition \ref{sw-over}, notice that the point $D^+$, which is obtained 
following the trajectory $\ga^+$ for
a time $\pi$ (see Figure \ref{f-sintesidoppia}), belongs to two different families of bang-bang 
trajectories at time $\pi$, one 
given by trajectories starting with control $-1$ and switching at time $s\leq t_1$, 
the other one given by trajectories that start with control $1$ and 
switching at time 
$s\in[\pi,t_3]$. Moreover, since $v(0)=\pi$, there must be a switching curve starting 
at $D^+$ and therefore we deduce that there are two possible behaviors of the 
optimal synthesis around this point: either this switching curve is optimal
or the two fronts continue to intersect generating an overlap curve.\\
Observe that if $\al\geq \pi/3$ the trajectories of the type $B_s B_{v(s)} B_t$
with $s$ small cannot be optimal since the vector fields $X^+_S$ and $X^-_S$ point
to opposite sides on the switching curve (i.e. the switching curve ``reflects the trajectories'', see 
footnote \ref{foot-piede}). In this case the two families of bang-bang trajectories described 
above must intersect giving rise to an overlap curve. Therefore to prove Proposition \ref{sw-over} we 
assume $\al<\pi/3$.\\\\
{\bf Proof of Proposition \ref{sw-over}}
First we parameterize the switching curve with respect to the first switching time
(assuming without loss of generality that this curve starts with $u=-1$): 
$$C(s)=e^{v(s)X^+_S} e^{sX^-_S}P_N\,.$$
We consider the functions $\xi_1(s)=\mbox{det}\,(C(s),C'(s),X^+_S (C(s)))$ (here the superscript $'$ 
denotes the 
derivative with respect to $s$)
and $\xi_2(s)=\mbox{det}\,(C(s),C'(s),X^-_S (C(s)))\,$.
It is easy to see that the optimality of $C(.)$, for $s$  small, depends 
on the signs of such
functions. Indeed $C(.)$ is locally optimal near the point $D^+=C(0)$ if 
and only if
for every $s>0$ (small enough) and given a neighborhood of $C(s)$ which is divided in two 
connected components $U_1,U_2$ by the  
trajectory $C(.)$, both $X^-_S (C(s))$ and $X^+_S (C(s))$ point towards $U_1$ or towards $U_2$.
It is easy to see that this happens if $\xi_1(s)$ and $\xi_2(s)$ have the same sign.
Notice that $\xi_1(0)=\xi_2(0)=0$ and that 
$\xi_1(s)=\mbox{det}\,(P_N,X^-_S (P_N),e^{-sX^-_S} X^+_S (e^{sX^-_S}P_N))=2\cos{\al}\sin^2\al\sin s$,
which is positive for every $\al<\pi/2$ and $s\in ]0,\pi[$. To determine the sign 
of $\xi_2(s)$ near $0$ it is enough to look at the sign of the derivative $\xi_2'(0)$
which can be computed directly: $\xi_2'(0)=4\cos\al\sin^2\al (1-2\sin^4\al)$.
We deduce that, if $\al<\arcsin(1/\sqrt[4]{2})$, the switching curve $C(.)$ is optimal for
$s$ small enough. For the particular value $\al=\arcsin(1/\sqrt[4]{2})$
one can easily check that the function $\xi_2(.)$
is negative for $s>0$ small, and then $C(.)$ is no more optimal for $\al\geq\arcsin(1/\sqrt[4]{2})$. 
The tangency of the switching curve starting at $D^+$ if $\al>\pi/4$,
is a consequence  of the fact that, in this case, the  bang-bang 
trajectory switching  at $D^+$ is an abnormal extremal (see Proposition 
\ref{l-ab} and \cite{libro}, Proposition 23 pag. 177). 
\quadp
\subsection{Time optimal trajectories reaching the south pole for $\al<\pi/4$} 
\llabel{a-pifferone2}
\ffoot{Forse e` meglio riferirsi a traiettorie estremali e non ottimali}
The purpose of this section is to characterize the optimal trajectories steering $P_N$ to $P_S$ in the case 
$\al<\pi/4$, i.e. to prove Proposition \ref{p-1} and \ref{p-2}. A key tool is Lemma \ref{v=v}.
Recall the shape of the function $v(s)$, in the case  $\al<\pi/4$ (see Figure \ref{f-vfg} A).  
Given $\al<\pi/4$ and $s\in [0,\pi]$ with $s\neq\arccos (-\tan^2 
\al)$, there exists one and only one time $s^\ast(s)\in[0,\pi]$ different from 
$s$, such that $v(s)=v(s^\ast(s))$. From Section \ref{s-traj-polo} recall the following definition of candidate optimal 
trajectories:\\
\phantom{~~~}$\bullet$~~$s_f=s^\ast(s_i)$ (i.e. TYPE-1-candidate optimal trajectories),\\
\phantom{~~~}$\bullet$~~$s_f=s_i$ (i.e. TYPE-2-candidate optimal trajectories)\\
A useful relation between  $s$ and $s^\ast(s)$ is given by the following:
\bl 
\llabel{l-belin}
For $\al<\pi/4$ and $s\in [0,\pi]$, it holds $s+s^\ast(s)= v(s)$. 
\el
{\bf Proof of Lemma \ref{l-belin}} 
Both $s$ and $s^\ast(s)$ satisfy the following equation in $t\in[0,\pi]$:	
$$
\cot\left(\frac{1}{2}v(s)\right)=-\frac{\sin (t)}{\cos 
(t)+\cot^2(\al)}
\Rightarrow\quad \cos\left(\frac{1}{2}v(s)-t\right)=-\cos
\left(\frac{1}{2}v(s)\right)\cot^2(\al) .
$$
Therefore, since 
$\,\ \frac{1}{2}v(s)-t\in[-\pi ,\pi ]\quad\forall s,t\in[0,\pi]\,$ and 
$s^\ast(s)\neq s$, it must be: 
$s^\ast(s)-\frac{1}{2}v(s)=\frac{1}{2}v(s)-s\quad\Rightarrow\quad
s+s^\ast(s)=v(s).$ \quadp\\\\
The description of candidate
optimal trajectories is simplified by 
the following lemma, of which we skip the proof.
\bl
Set: $$Z(s)=\frac{1}{\rho}
\left(\begin{array}{ccc}
0 & \cot\left(\frac{1}{2}v(s)\right) & -\sin(\al) \\
-\cot\left(\frac{1}{2}v(s)\right) & 0 & 0  \\
\sin(\al)  & 0 & 0 
\end{array}\right)
$$
where $~\rho=\sqrt{\cot^2\left(\frac{1}{2}v(s)\right)+
\sin^2(\al)}~$.
Then, if $\th(s)$ is defined as in \r{thth}, 
we have $e^{\th (s) 
Z(s)}=e^{v(s) X^-_S}e^{v(s) X^+_S}$.
\el
Notice that the matrix $\ Z(s)\in so(3)\ $ is normalized in such a way that 
the map $\ t\mapsto e^{t Z(s)}\in SO(3)\ $ represents a rotation around the 
axes  $R(s)=\big(0,\sin(\al),\cot (\frac{1}{2}v(s))\big)^T$ with angular 
velocity equal to one.

To prove the results stated in Section~\ref{s-traj-polo} we study separately 
the two possible cases listed above:\\\\
{\bf Proof of Proposition \ref{p-1}.}
In this case we consider TYPE-1-candidates optimal trajectories.
Assume that the optimal trajectory starts with $u=-1$ (the case 
$u=1$ is symmetric) and has an even number $n$ of switchings. Then it must be 
\bqn
P_S=e^{s_f X^-_S}\underbrace{e^{v(s_i)X^+_S}\!\!\!\!\!\!\!\ldots\ldots\ e^{v(s_i)
X^+_S}}_{n-1\  times} e^{s_i X^-_S}P_N
\eqnl{pigreco}
where $P_N$ and $P_S$ denote respectively the north and the south pole, 
and we 
have that
$$e^{s_i X^-_S}P_S=e^{v(s_i)X^-_S}e^{v(s_i)X^+_S}\!\!\!\!\!\!\!\ldots\ldots\  
e^{v(s_i)X^+_S} e^{s_i X^-_S}P_N=e^{\frac{1}{2}n\th(s_i)Z(s_i)} e^{s_i X^-_S}P_N$$
from which we deduce that $s_i$ must satisfy 
$$\frac{1}{2}n\th(s_i)=\pi+2p\pi\ \mbox{ for some integer}\ p.$$
It is easy to see that a value of $s_i$ which satisfies previous equation 
with $p>0$ doesn't give rise to a candidate optimal trajectory since the corresponding 
number of switchings is larger than $N_M$. Therefore in previous equation
it must be $p=0$. If $n$ is odd the relation \r{pigreco} becomes
\bqn
P_S=e^{s_f X^+_S}\underbrace{e^{v(s_i)X^-_S}\!\!\!\!\!\!\!\ldots\ldots\ e^{v(s_i)
X^+_S}}_{n-1\  times} e^{s_i X^-_S}P_N
\eqnl{pigreco2}
and, moreover, by symmetry:
$$P_N=e^{s_f X^-_S}e^{v(s_i)X^+_S}\!\!\!\!\!\!\!\ldots\ldots\  e^{v(s_i)X^-_S} 
e^{s_i X^+_S}P_S.$$
Then, combining with \r{pigreco2} and using the relation Lemma \ref{l-belin}, we find:
$$P_N=e^{-s_i X^-_S}\underbrace{e^{v(s_i)X^-_S}\!\!\!\!\!\!\!\ldots\ldots\  
e^{v(s_i)X^+_S}}_{2n\  times} e^{s_i X^-_S}P_N=e^{-s_i X^-_S}e^{n\th(s_i)Z(s_i)} e^{s_i X^-_S}P_N .$$
Since $e^{s_i X^-_S}P_N$ is orthogonal to the rotation axis $R(s_i)$
corresponding to $Z(s_i)$, previous identity is satisfied if and only if 
$n\th(s_i)=2m\pi$ with $m$ positive integer. As in the previous case, for
a candidate optimal trajectory, it must be $m=1$. \quadp \\\\
{\bf Proof of Proposition \ref{p-2}.}
Here we consider TYPE-2-candidate optimal trajectories.
For simplicity call $s_i=s_f=s$. Assume, as before, that the 
optimal trajectory starts with $u=-1$ . If this trajectory has $n=2q+1$ 
switchings then it must be $$P_S=e^{s X^+_S}e^{q\th(s)Z(s)} e^{sX^-_S}P_N.$$
In particular the points $e^{-sX^+_S} P_S$ and 
$e^{sX^-_S} P_N$ must belong to a plane invariant with respect to
rotations generated by $Z(s)$ and therefore the difference
$e^{sX^-_S} P_N-e^{-sX^+_S} P_S$ must be orthogonal to the rotation axis $R(s)$.
Actually it is easy to see that this is true for every value $s\in[0,\pi]$, 
since both $e^{-sX^+_S} P_S$ and $e^{sX^-_S} P_N$ are orthogonal to $R(s)$.
Moreover, since the integral curve of $Z(s)$ passing through $e^{sX^-_S} P_N$ 
and $e^{-sX^+_S} P_S$ is a circle of radius 1, it is easy to 
compute the angle $\beta(s)$ between these points. In particular the distance 
between $e^{sX^-_S} P_N$ and $e^{-sX^+_S} P_S$ coincides with $2 \sin (\frac{\beta
(s)}{2})$ , and so one easily gets the expression
$\beta(s)=2 \arccos(\sin(\al)\cos(\al) (1-\cos(s))).$
Then Proposition~\ref{p-2} is proved when $n$ is odd.

Assume now that the optimal trajectory has $n=2q+2$ switchings, then we can 
assume without loss of generality that
$P_S=e^{sX^-_S}e^{v(s)X^+_S}e^{q\th(s)Z(s)} e^{sX^-_S}P_N\ .$
First of all it is possible to see that $e^{-v(s)X^+_S}e^{-sX^-_S}P_S$ is 
orthogonal to $R(s)$. 
So it remains to compute the angle $\tilde{\beta}(s)$ between the point 
$\,e^{sX^-_S}P_N\,$ and the point $\,e^{-v(s)X^+_S}e^{-sX^-_S}P_S\,$ on the plane 
orthogonal to $R(s)$. As before the distance between these points coincides 
with $2 \sin (\frac{\tilde{\beta} (s)}{2})$.
Instead of computing directly $\tilde{\beta}(s)$ we compute the difference 
between the angles $\tilde{\beta}(s)$ and the angle $\beta(s)$.
We know that
$$
2 \sin (\frac{\beta(s)}{2}-\tilde{\beta}(s))=|e^{-v(s)X^+_S}e^{-sX^-_S}P_S-
e^{-sX^+_S} P_S|=|e^{-sX^-_S}P_S-e^{v(s)X^+_S}e^{-sX^+_S} P_S|=|e^{-sX^-_S}P_S-e^{s^\ast(s)X^+_S}P_S|. $$
Using the fact that $s$ and $s^\ast(s)$ satisfy the relation $v(s)=v(s^\ast(s))$ one 
can easily find that $$|e^{-sX^-_S}P_S-e^{s^\ast(s)X^+_S}P_S|=2\sqrt{1-\cos^2(\al)\sin^2 
\left(\frac{1}{2}v(s)\right)}.$$
Therefore $\beta(s)=\tilde{\beta}(s)+2 \arccos{\left(\cos(\al)\sin\left(
\frac{1}{2}v(s)\right)\right)}.$
This leads to $\beta(s)-\tilde{\beta}(s)=\th(s)/2$ and the 
proposition is proved also in the case $n$ is even.\quadp
\subsubsection{Proof of Proposition~\ref{alterna}, 
on the alternating behaviour of the optimal synthesis} 
\llabel{s-alterna}
In this section we need to consider also the dependence on $\al$ of the functions 
$v(s),\th(s),\beta(s),\mathcal{F}(s),\mathcal{G}(s)$. Therefore we switch to the notation 
$v(s,\al),\th(s,\al),\beta(s,\al),\mathcal{F}(s,\al),\mathcal{G}(s,\al)$.

The claims on existence of solutions of Proposition~\ref{alterna} come 
from the fact that  $\mathcal{F}(0)=\mathcal{F}(\pi)=\frac{\pi}{2 \al}$ and the 
only minimum point of $\mathcal{F}$ occurs at $\bar{s}=\pi-\arccos(\tan^2(\al))$.
It turns out that the image of $\mathcal{F}$ is a small interval whose
length is of order $\al$ and therefore equation \r{effe} has a 
solution only if $\al$ is close enough to $\frac{\pi}{2m}$ for some integer 
number $m$. This proves {\bf C.} i.e. the existence of $r_2(m)$ satisfying 
$r_2(m)=O(1/m)$.  

On the other hand it is possible to estimate the derivative of $\mathcal{G}$ 
with respect to $s$ showing that it is negative in the open interval 
$]0,\pi[$.
Therefore, since $\mathcal{G}(0)=\frac{\pi}{2 \al}+1$ and 
$\mathcal{G}(\pi)=\frac{\pi}{2 \al}-1$, equation \r{gi} has always two 
positive solutions. 

For the particular values $\al=\frac{\pi}{2m}$, where $m>1$ is an integer number, the solutions 
to the equations \r{effe} and \r{gi} give rise
to two candidate optimal trajectories, one with $m$ bang arcs, all of length $\pi$
(TYPE-1 and TYPE-2 candidate optimal trajectory at the same time), while the second one has 
one more switching and is a TYPE-2 candidate optimal trajectory.
We want to see that the optimal trajectory is the first one. To this purpose, we need to estimate 
the time needed to reach the south pole by the second candidate 
optimal trajectory showing that it is greater than $m\pi=\frac{\pi^2}{2\al}$.

First, using the Taylor expansions with respect to $\al$ and centered at $0$ of $\beta(\pi/2,\al)$ 
and $\th(\pi/2,\al)$, one obtains
\bqn
\mathcal{G}(\frac{\pi}{2},\al)=\frac{\pi}{2\al}-\al\frac{\pi}{4}+o(\al). 
\eqnl{espansione}
We want now to estimate the solution $s(\al)$ of the equation $\mathcal{G}(s,\al)=\frac{\pi}{2\al}$.
This can be done using the previous expression \r{espansione} and since it is possible to 
estimate the derivative of $\mathcal{G}(.)$ with respect to $s$ near $s=\pi/2$, in the 
following way:
$$\frac{{\rm d}}{{\rm d}s}\mathcal{G}(s,\al)=-1+o\left(|\al|+|\frac{\pi}{2}-s|\right),$$
then it is easy to find that $s(\al)=\displaystyle{\frac{\pi}{2}-\al\frac{\pi}{4}+o(\al)}$, and,
consequently $v(s(\al),\al)=\pi+2\al^2+o(\al^2)$.
Therefore 
$\displaystyle{2s(\al)+\left(\frac{\pi}{2\al}-1\right)v(s(\al),\al)=\frac{\pi^2}{2\al}+\al\frac{\pi}{2}
+o(\al)}$. In particular, for $\al=\frac{\pi}{2m}$ this expression coincides with the time needed 
to reach the south pole by the candidate optimal trajectory and, since for $m$ large enough it is 
larger than $m\pi=\frac{\pi^2}{2\al}$, we found that this trajectory cannot be optimal.
Since the solutions to the equations \r{effe},\r{gi} change continuously with respect to $\al$
for each fixed number of switchings $n$, we easily deduce that if we slightly decrease $\al$ 
starting from the value $\frac{\pi}{2m}$ the solution of \r{effe} for $n=m$ does not give rise 
to an optimal trajectory.

For $\al$ slightly smaller than $\bar{\al}:=\frac{\pi}{2m}$ there is a TYPE-2 candidate optimal
trajectory corresponding to a solution $(s_1(\al),m+1)$ of \r{gi}, where $s_1(.)$ is continuous
(on $[\bar \al-\eps,\bar \al]$) and $s_1(\bar{\al})=0$, and there is also a TYPE-1 candidate optimal 
trajectory corresponding to a solution $(s_2(\al),m)$ of \r{effe} where $s_2(.)$ 
is continuous (on $[\bar \al-\eps,\bar \al]$) and $s_2(\bar{\al})=0$. Clearly for $\al=\bar{\al}$ 
these trajectories coincide. So we have to compare the time to reach the south pole for such 
trajectories with $\al$ close to $\bar{\al}$.

We start with the TYPE-1 candidate optimal trajectory. From equation \r{effe} we have that 
$\displaystyle{\frac{{\rm d}}{{\rm d}\al}\th(s_2(\al),\al)=0}$. 
We use a subscript $s$, $\al$ to denote the partial differentiation with respect to such variables.
Since $\th_s(0,\al)=0$ we cannot apply directly the implicit function theorem near
$(0,\bar{\al})$. However, if we set $\tilde{s}_2(\al)=s_2^2(\al)$ we find that 
$\displaystyle{\tilde{s}_2'(\al)=\frac{2s_2(\al)\th_{\al}(s_2(\al),\al)}{\th_s(s_2(\al),\al)}}$  
(the superscript $'$ denotes differentiation with respect to $\al$), and then, passing 
to the limit as $(s_2(\al),\al)$ tends to $(0,\bar \al)$, one 
easily finds that $\displaystyle{\tilde{s}_2'(\bar\al)=-\frac{2}{\sin(\bar\al)^3 \cos(\bar\al)}}\,$. 

Now we want to determine the way in which the total time $T_2(\al)=m v(s_2(\al),\al)$ changes. 
It is easy to see that $T_2(\al)$ is not differentiable at $\bar \al$, therefore we introduce 
the function $F(\al)=(T_2(\al)-T_2(\bar\al))^2=m^2(v(s_2(\al),\al)-\pi)^2$.

Then $\displaystyle{\frac{{\rm d}}{{\rm d}\al} F(\al)=2m^2\frac{{\rm d}}{{\rm d}\al}v\big(s_2(\al),\al\big)
\big(v\big(s_2(\al),\al\big)-\pi\big)=2m^2 \Big(v_s\big(s_2(\al),\al\big)s_2'(\al)+ 
v_{\al}\big(s_2(\al),\al\big)\Big) }$ and, after the substitution $\displaystyle{s_2'(\al)=\frac{\tilde{s}_2'(\al)}{2s_2(\al)}}$ we can pass to the limit as $\al$ converges to $\bar\al$
obtaining 
$$\frac{{\rm d}}{{\rm d}\al} F(\al)|_{\al=\bar\al}=m^2 v_s^2(0,\bar \al)\tilde{s}_2'(\bar\al)=
-8 m^2 \tan{\bar\al}\,.$$

Now we consider the TYPE-2 candidate optimal trajectory and we want to estimate $s_1(\al)$.
From equation \r{gi} we have that $s_1(.)$ is implicitly defined by the equation 
$\Phi(s_1(\al),\al):=2\beta(s_1(\al),\al)-m \th(s_1(\al),\al)=0$. As before it is easy to see that $s_1(.)$ 
is not differentiable at $\bar\al$ and therefore we introduce the parameter $\tilde s_1(\al)=s_1^2(\al)$.
As before, it is possible to compute the derivative $\tilde s_1'(\al)$:
$$\tilde s_1'(\bar\al)=-\lim_{\al\to\bar\al}
\frac{2 s_1(\al)\Phi_{\al}(s_1(\al),\al)}{\Phi_s(s_1(\al),\al)}=
-\frac{2m}{\sin\bar\al\cos\bar\al (1+m\sin^2\bar\al)}\,.$$
We have now to estimate the total time $T_1(\al)=2 s_2(\al)+m v\big(s_2(\al),\al\big)$ for $\al$ close
to $\bar\al$. Define $$G(\al)=\big(T_1(\al)-T_1(\bar\al)\big)^2=
\Big(2s_2(\al)+m\big(v(s_2(\al),\al)-\pi\big)\Big)^2\,,$$
then we compute the derivative of $G(.)$ as follows:
$$\frac{{\rm d}}{{\rm d}\al} G(\al)|_{\al=\bar\al}=\lim_{\al\to \bar\al} \Bigg[2\Big(2s_2(\al)+
m\big(v\big(s_2(\al),\al\big)-\pi\big)\Big)\Big(\frac{\tilde s_2'(\al)}{s_2(\al)}+
m\Big(\frac{v_s(s_2(\al),\al)\tilde s_2'(\al)}{2 s_2(\al)}+v_\al \big(s_2(\al),\al\big)\Big)\Big)\Bigg]=
$$
$$=\lim_{\al\to \bar\al}\Bigg[2\Big(2+m\frac{v\big(s_2(\al),\al\big)-v\big(0,\al\big)}{s_2(\al)}
\Big)\Bigg]\lim_{\al\to \bar\al}\Bigg[\tilde s_2'(\al)+
m\Big(\frac{1}{2} v_s(s_2(\al),\al)\tilde s_2'(\al)+v_\al \big(s_2(\al),\al\big)s_2(\al)\Big)\Bigg]=$$
$$=\big(2+mv_s(0,\bar\al)\big)^2s_2'(\bar\al)=-\big(2+2m\sin^2\bar\al\big)^2
\frac{2m}{\sin\bar\al\cos\bar\al (1+m\sin^2\bar\al)}=-\frac{8m(1+m\sin^2\bar\al)}
{\sin\bar\al\cos\bar\al}\,.$$
Since $$\frac{8m(1+m\sin^2\bar\al)}{\sin\bar\al\cos\bar\al}>m\tan\bar \al$$ we deduce that
$G(\al)$ decreases faster than $F(\al)$ as $\al$ goes to $\bar\al$ and, since $T_1(\al)$ and
$T_2(\al)$ are decreasing for $\al$ close to $\bar\al$, we have that $T_2(\al)>T_1(\al)$, i.e.
the TYPE-1 trajectory is optimal for $\al\in [\bar\al-\eps,\bar\al]$.

\section{The Time Needed to Reach 
Every Point of the Bloch Sphere Starting 
from the North Pole in the case
 $\al\in[\pi/4,\pi/2[$}
\llabel{a-c}

In this section we assume  $\al\in[\pi/4,\pi/2[$.
If $\al$ is close to $\pi/4$ it is easy to verify that the south 
pole is not the last point reached by bang-bang trajectories 
(the last point reached belongs to the cut locus present in the region $\ON^\pm$) 
and the time needed to cover the whole sphere is slightly larger than $2\pi$.

On the other hand, if $\al$ is large enough then the velocity along a 
singular arc is small and therefore the time needed to move along trajectories
containing singular arcs is larger than $2\pi$.
The following proposition gives the asymptotic behaviour of the total
time needed to reach every point from the north pole and determines the last 
point reached by the optimal synthesis for $\al$ large enough. 
\bp 
\llabel{p-ultima}
Let $T(\al)$ the time needed to cover the whole sphere. Then, if $\al$ is
large enough
\bqn
T(\al)=\frac{\pi}{2\cos{\al}}+\pi-\frac{2\arcsin(\cot{\al})}{\cos{\al}}+2\arcsin(\cot^2\al)= 
\frac{\pi}{2\cos{\al}}+\pi-2+O\Big(\frac{\pi}{2}-\al\Big)
\eqnl{tempo}
and the last points reached for a fixed value of $\al$ are 
$\pm(\sqrt{1-\cot^2\al},\cot\al,0)^T$.
\ep
{\bf Proof of Proposition \ref{p-ultima}}
From Proposition \ref{propA} the last points reached by optimal trajectories of the form
$B_t S_s B_{t'}$ must lie on overlap curves which are subsets of the equator. 
Therefore it is enough to estimate the maximum time to reach these overlap curves. 
Assume that the first bang arc corresponds to the control $u=1$ 
and denote by $\beta$ the angle corresponding to the arc of the equator between 
the last point of the singular arc and the point $O^+=(1,0,0)^T$. Notice that $\beta\in]0,\arccos(\cot\al)[$. 
Then it is easy to
find the expression $T(\al,\beta)$ of the time needed to reach the overlap curve 
along that optimal trajectory:
$$T(\al,\beta)=\pi-\arccos(\cot^2\al)+\frac{\arccos(\cot\al)}{\cos{\al}}-\frac{\beta}{\cos{\al}}+
\arccos\left(\frac{\cos^2\al-\tan^2\beta}{\cos^2\al+\tan^2\beta}\right)\,.$$
The conclusion follows finding the maximum with respect to $\beta$ of the previous 
quantity, which corresponds to the value $\bar{\beta}=\arcsin(\cot\al)$.
Notice that  $\bar{\beta}$ belongs to the interval of definition of $\beta$ only if  
$\al>\mathrm{arccot}(\sqrt{2}/2)$.
\quadp
\brem
If $\al>\mathrm{arccot}(\sqrt{2}/2)$
there are two symmetric neighborhoods of the points  $\pm(\sqrt{1-\cot^2\al},\cot\al,0)^T$ 
that are not reached by optimal trajectories at time $t<T(\al)$ with $t$ close enough to 
$T(\al)$, i.e. the reachable set is not simply connected.
\ffoot{nota che la topologia cambia cosi`:
$R^2$(top. equivalente all'intorno di N) --$>R^2$-due punti=$S^2$-3pti (--$>R^2$ per alfa non 
troppo grande oppure --$>S^2$-2 pti=$R\times S$ per alfa grande)--$>S^2$}
\erem
\brem
Recall that for system \r{eq-se} the time needed to cover the whole sphere for $\al$ 
close enough to $\pi/2$ is obtained dividing by $k=\frac{2E}{\cos{\al}}$ the expression 
\eqref{tempo}. Therefore, if we fix $E$ it turns out that this quantity converges to 
$\frac{\pi}{4E}$ as $M$ goes to infinity.
\erem

\medskip



\end{document}